\documentclass[a4paper,twocolumn,11pt, accepted=2024-07-11]{quantumarticle}

\pdfoutput=1
\usepackage[utf8]{inputenc}

\usepackage{bbm}
\usepackage{amsmath}
\PassOptionsToPackage{hyphens}{url}
\usepackage{float}
\usepackage[hidelinks]{hyperref}
\usepackage{graphicx}
\usepackage{graphics}
\usepackage{amssymb}
\usepackage{mathtools}
\usepackage{xcolor}
\usepackage{physics}
\usepackage[british]{babel}
\usepackage{csquotes}
\usepackage{graphicx}
\usepackage{qcircuit}
\usepackage{enumitem}
\setlist{nosep} 
\usepackage{stmaryrd}
\usepackage{algorithm}
\usepackage{algorithmic}
\usepackage{multirow}
\usepackage{qcircuit}
\usepackage{changes}
\usepackage[numbers,sort&compress]{natbib}

\newlength\myheight
\newlength\mydepth
\settototalheight\myheight{Xygp}
\newcommand*\inlinegraphics[1]{%
  \settototalheight\myheight{Xygp}%
  \settodepth\mydepth{Xygp}%
  \raisebox{-\mydepth}{\includegraphics[height=\myheight]{#1}}%
}

\floatstyle{ruled}

\newfloat{protocol}{htb!}{idf}
\floatname{protocol}{Protocol}

\newfloat{resource}{htb!}{idf}
\floatname{resource}{Resource}

\newfloat{simulator}{htb!}{idf}
\floatname{simulator}{Simulator}

\usepackage{amsthm}

\definecolor{dark-gray}{gray}{0.40}
\definecolor{quantumviolet}{HTML}{53257F} 
\definecolor{quantumlightviolet}{HTML}{A088B1}
\definecolor{quantumgreen}{HTML}{00826F}
\definecolor{quantumrose}{HTML}{EDB3FF} 
\definecolor{quantumdarkrose}{HTML}{F06292}
\definecolor{quantumturquoise}{HTML}{00C9AF}
\definecolor{quantumblue}{HTML}{85B1CC}
\definecolor{quantumdarkgray}{HTML}{4C4452}
\definecolor{quantumgray}{HTML}{555555}

\begin{document}

\title{A Spin-Optical Quantum Computing Architecture}
\author{Grégoire de Gliniasty}
\affiliation{Quandela, 7 Rue Léonard de Vinci, 91300 Massy, France}
\affiliation{Sorbonne Université, CNRS, LIP6, F-75005 Paris, France}

\author{Paul Hilaire}
\email{paul.hilaire@quandela.com}
\affiliation{Quandela, 7 Rue Léonard de Vinci, 91300 Massy, France}

\author{Pierre-Emmanuel Emeriau}
\affiliation{Quandela, 7 Rue Léonard de Vinci, 91300 Massy, France}
\author{Stephen C. Wein}
\affiliation{Quandela, 7 Rue Léonard de Vinci, 91300 Massy, France}
\author{Alexia Salavrakos}
\affiliation{Quandela, 7 Rue Léonard de Vinci, 91300 Massy, France}
\author{Shane Mansfield}
\affiliation{Quandela, 7 Rue Léonard de Vinci, 91300 Massy, France}

\begin{abstract}
\noindent We introduce an adaptable and modular hybrid architecture designed for fault-tolerant quantum computing. 
It combines quantum emitters and linear-optical entangling gates
to leverage the strength of both matter-based and photonic-based approaches.
A key feature of the architecture is its practicality, grounded in the utilisation of experimentally proven optical components. 
Our framework enables the execution of any quantum error correcting code, but in particular maintains scalability for low-density parity check codes by exploiting built-in non-local connectivity through distant optical links.
To gauge its efficiency, we evaluated the architecture using a physically motivated error model. 
It exhibits loss tolerance comparable to existing all-photonic architecture but without the need for intricate linear-optical resource-state-generation modules that conventionally rely on resource-intensive multiplexing.
The versatility of the architecture also offers uncharted avenues for further advancing performance standards.

\end{abstract}

\maketitle
\section{Introduction}

Fault-tolerant (FT) quantum computing (QC) allows arbitrary quantum algorithms to be performed even in the presence of moderated yet non-negligible noise, thanks to the threshold theorem~\cite{aharonov1997fault, knill1998resilient, kitaev2003fault}. 
However appealing this concept might be in theory, the practical realization of an FT quantum computer is a highly nontrivial challenge.

A QC architecture is the association of two main ingredients.
The first is a theoretical model for quantum computing
such as gate-based~\cite{barenco1995elementary}, measurement-based~\cite{raussendorf2001one}, ancilla-driven~\cite{anders2010ancilla} or adiabatic~\cite{das2008colloquium} quantum computing. The second is the organization of the hardware components that enables the physical implementation of this quantum computing model. Most proposals~\cite{van2010distributed, monroe2014large, nickerson2014freely, bombin2021interleaving, bourassa2021blueprint, chamberland2022building, bartolucci2023fusion} 
for the physical implementation of FTQC rely on a modular approach.
The concept of a \emph{scalable} modular FT quantum computer revolves around a fundamental principle: while various module types may exist, the number of module categories, their specifications, and their quality should remain constant as the quantum computer grows in size.
To ensure scalability, it is thus essential that the noise level of each hardware module remains independent of the quantum computer's size, thereby preventing it from straying out of the fault-tolerant regime.

While it may seem self-evident, it is essential to underscore that an FTQC architecture designed for superconducting qubits will inherently differ from one tailored to linear optical systems which for instance lacks deterministic two-qubit gates.
Crucially, an efficient FTQC architecture cannot be universally ``hardware-agnostic''. It demands thorough adaptation to the unique characteristics of a particular platform and its corresponding noise model.

In the realm of photonic implementation, Ref.~\cite{bartolucci2023fusion}
proposed a modular photonic FTQC architecture, which places specific emphasis on the utilization of spontaneous parametric down-conversion (SPDC) sources in combination with linear-optical interferometers and detectors.
However, these sources operate through a heralded generation process characterized by low probability, and entanglement is generated probabilistically  using linear-optical gates.

On the other hand, in recent years we have witnessed the emergence of highly efficient single-photon sources, driven by a fundamentally distinct technology paradigm, specifically, the utilization of individual atoms or artificial atoms.
Although SPDC sources had previously held the record for creating the largest photonic entangled states, atom-based single-photon sources have since surpassed them by generating a 14-qubit GHZ state~\cite{thomas2022efficient}.
These sources exploit the intricate interplay of light and matter, where the spin of the quantum emitter is coupled to the polarization of the emitted photon. Precise control over this spin enables the growth of photonic graph states.
Contrary to SPDC sources, the quantum emitter's spin mediates the generation of entanglement between photons which becomes a deterministic process.
Moreover, single-photon sources based on trapped ions~\cite{blinov2004observation} or artificial atoms such as semiconductor quantum dots can exploit the same strategy to efficiently generate photonic graph states.
The latter has already allowed experimental demonstration of photonic graph state generation~\cite{schwartz2016deterministic, coste2023high, cogan2023deterministic, meng2023deterministic, meng2023photonic}, which it can achieve at rates that are orders of magnitude higher than atomic sources provided the photon collection efficiency is increased.

While quantum-emitter-based sources can replace SPDC sources to implement the architecture of Ref.~\cite{bartolucci2023fusion} with a smaller footprint~\cite{hilaire2022near}, it remains uncertain whether that scheme is optimally suited for these particular platforms. Indeed, these sources use a spin degree of freedom as a ``photon entangler'' to produce photonic entangled states~\cite{schon2005sequential, lindner2009proposal}, which are potentially useful for a fully-photonic FTQC \cite{,paesani2022high,herrera2010photonic}. 
Yet, relying solely on the quantum emitter as a source of entangled photons might not be the optimal strategy either, as there is potential to harness its spin qubit to encode and process quantum information, thus maximizing its utility.
What we need is a hybrid architecture combining the best of 
both worlds, by exploiting both the spin and photonic qubits to their full potential.

In this paper, we introduce a modular and scalable FTQC architecture tailored for quantum-emitter-based platforms. Our study delves into the investigation of its performance characteristics with a precise characterization of the thresholds for quantum error correction in the presence of a realistic noise model. 
In this framework, the physical qubits are encoded in the quantum emitters, while the photons are used as ancilla qubits. We use these ancilla photons to implement nearly-deterministic gates between different emitters, employing a repeat-until-success (RUS) linear-optical scheme~\cite{bose1999proposal, lim2005repeat}. We also investigate a variant called hybrid RUS gates to boost the tolerance with respect to coherence time of the spin at the expense of loss tolerance. 

The paper is organized as follows.
In Sec.~\ref{sec_spoqc}, we detail the hardware layout of our quantum computing architecture.
In Sec.~\ref{sec_rus_gate}, we focus on the linear-optical RUS gate on which our architecture heavily relies to generate two-qubit entanglement.
We compare our architecture to others in Sec.~\ref{sec_comp}.
We then present in Sec.~\ref{sec_results} the performance in terms of FT thresholds of our architecture, based on a physically relevant error model.
We finally discuss our results, compare them with the literature in Sec.~\ref{sec_comp}, and conclude in Sec.~\ref{sec_discussion}.

\begin{figure*}[!ht]
  \centering
  \includegraphics[width=16cm]{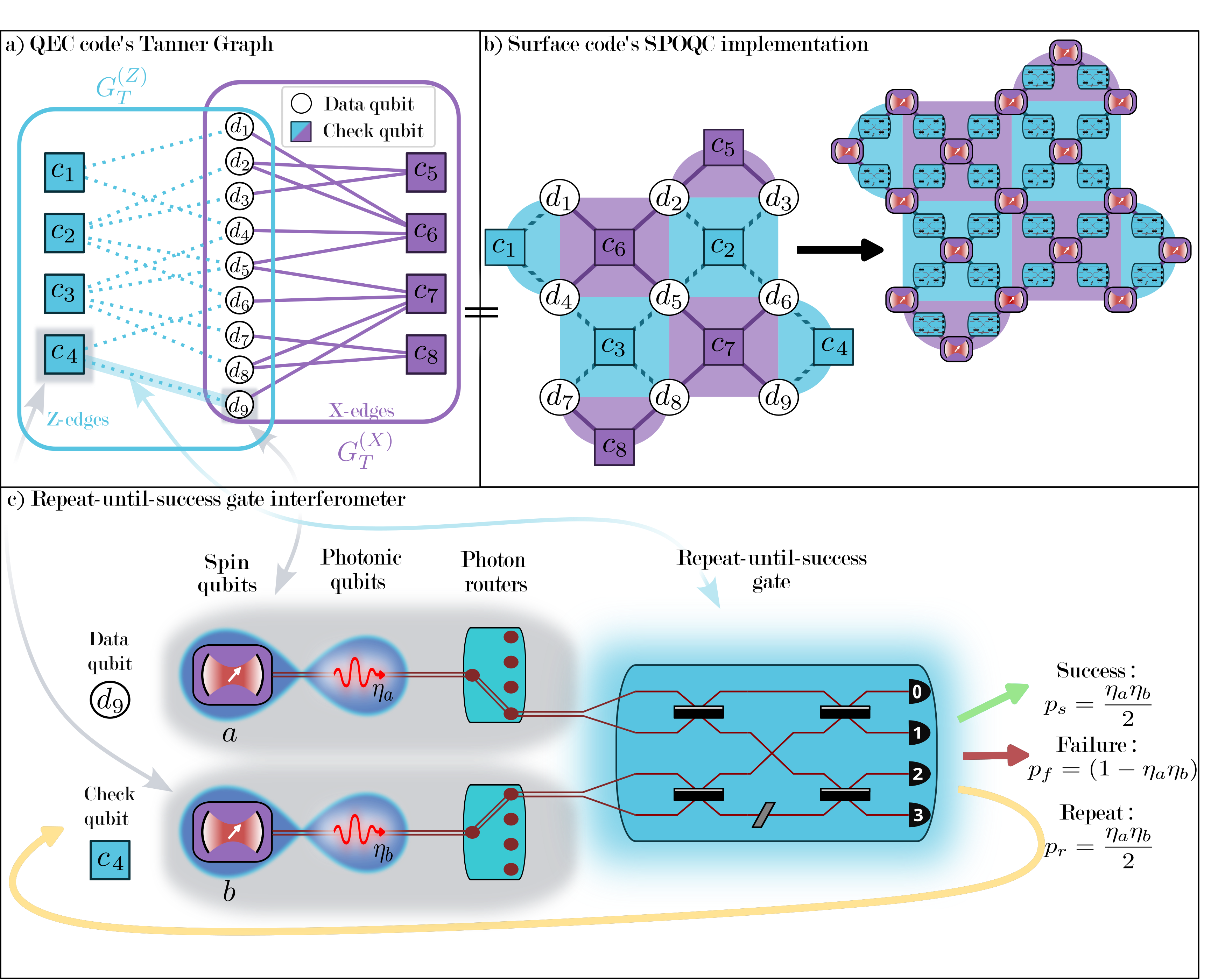}
  \caption{
  (a) Tanner graph of a $3 \times 3$ rotated surface code with its subgraphs $G_T^{(X)}$ and $G_T^{(Z)}$. $X$-edges (solid, violet) and $Z$-edges (dashed, blue) belong respectively to $E_T^{(X)}$, and $E_T^{(Z)}$.
  (b) Macroscopic architecture for a rotated surface code.
  Left: The previous Tanner graph represented in a planar layout.
  Right: The SPOQC architecture for the $3 \times 3$ rotated surface code.
  (c) Linear-optical repeat-until-success gate including spin qubits capable of emitting entangled photons, coupled to photon routers, an interferometer and detectors.
  The linear-optical interferometer (produced using the \emph{Perceval}~\cite{heurtel2023perceval} framework) performs the unitary from Eq.~\eqref{eq_unitary_rus}.
  Horizontal black rectangles correspond to $50:50$ beamsplitters, the inclined parallelogram corresponds to a $-\pi/2$ phase shifter, numbered semiovals are photon-number-resolving detectors.}
  \label{fig_full}
\end{figure*}

\section{Spin-optical quantum computing architecture} \label{sec_spoqc}

In the following, we present our proposal for an FTQC architecture, hereafter referred to as the spin-optical quantum computing (SPOQC) architecture. 
This approach adopts a modular design specifically tailored for quantum emitter platforms. In this architecture, each quantum emitter is assumed to embed a qubit degree of freedom -- its spin -- and is capable of emitting spin-entangled photonic qubits.
Crucially, we do not require direct interaction between quantum emitters, which makes this approach suitable even for isolated quantum emitters. 
Instead, we use photons to mediate the two-qubit gates between the spins of non-interacting quantum emitters. 
This is achieved through linear-optical gates, in particular RUS gates, which will be discussed in greater detail later in this paper.

\subsection{Global overview of the architecture}

We first provide a global overview of the SPOQC architecture, illustrated by Fig.~\ref{fig_full}. Importantly, this architecture is compatible with any quantum error correcting (QEC) code. 
In particular, it is 
compatible with 
quantum low-density parity check (LDPC) codes~\cite{kitaev2003fault, bombin2006topological, breuckmann2021quantum} and can implement codes with \emph{non-local connectivity}. 
This important category of codes encompasses many well-known QEC codes, including the well-established surface codes~\cite{kitaev2003fault} but also the recently discovered ``good'' LDPC codes~\cite{leverrier2022quantum, panteleev2022asymptotically}, i.e.\ LDPC codes with constant encoding rate and linear distance.

To enable the fault-tolerance capabilities of a QEC code, efficient error detection is essential, requiring the use of dedicated ``check'' qubits.
In the SPOQC architecture,  
each data or check qubit corresponds to a spin qubit within a quantum emitter, and their arrangement mirrors the connectivity defined by the QEC code's Tanner graph. This graph serves as a useful visual representation of the QEC code's structure.
In this graph, each edge corresponds to optical links, such as fibers, a linear-optical interferometer, and photon detectors. This setup enables the execution of linear-optical gates between photons from different quantum emitters, ultimately resulting in the implementation of a two-qubit spin entangling gate.
Crucially, if a spin node in the graph has more than one neighbor (which is usually the case), it requires an optical router to orient the photons to the correct photonic link.

For quantum LDPC codes~\footnote{see Appendix~\ref{subsec_nonlocal_ldpc} for a more detailed discussion on the scalability for non-local LDPC code.}, each node of the Tanner graph has a bounded number of neighbors which is critical for the scalability of the architecture. 
Without this constraint, the number of output modes of some photon routers could potentially become unbounded as the size of the QEC code increases.
The architecture remains inherently fault-tolerant as long as the FT gates require a bounded amount of additional connectivity between the physical qubits. This is for example the case for surface code implementations with lattice surgery and magic state distillation~\cite{litinski2019game}.

Fig.~\ref{fig_full} summarizes the essential features of the SPOQC architecture, taking the $3 \times 3$ rotated surface code as an 
example.
We use the Tanner graph of the QEC code (Fig.~\ref{fig_full}(a)) to design the macroscopic physical layout of the architecture (Fig.~\ref{fig_full}(b)), based on simple optical modules.
The connectivity of this layout enables the indirect measurement of stabilizers using check qubits.
Physically, each qubit is encoded in a quantum emitter's spin and the entangling gates are implemented optically using the linear-optical interferometer depicted in Fig.~\ref{fig_full}(c).
In the following, we present this architecture in more detail.

\subsection{Tanner graph of a quantum error correcting code}

An FT quantum computer relies on quantum error correction to actively detect and correct errors.
Quantum logic gates can be implemented fault-tolerantly using different strategies depending on the QEC code being used, such as magic state distillation~\cite{bravyi2005universal}, lattice surgery~\cite{horsman2012surface}, or code switching~\cite{anderson2014fault}. 
In the following, we will not consider the different methods to implement FT gates but instead orient our discussion to focus on the robustness of our quantum computing architecture to errors. 
In accordance with the threshold theorem~\cite{aharonov1997fault, knill1998resilient, kitaev2003fault}, for a specific error model, any quantum algorithm can be executed fault-tolerantly, under the condition that the physical errors remain below a certain threshold value, and that the FT quantum computer is sufficiently large. This threshold value depends on the chosen methods for implementing error correction and the specific QEC code employed.

In an $\llbracket n,k,d \rrbracket$ QEC code, $n$ data qubits encode $k < n$ logical qubits to protect them from at least $t = \lfloor (d-1)/2 \rfloor$ qubit errors (or $d-1$ qubit losses). Errors are identified by measuring stabilizer check operators which depend on the QEC code being used. 

In an $\llbracket n,k,d \rrbracket$ stabilizer QEC code, there are $(n-k)$ independent stabilizer check operators, which are multi-qubit Pauli operators acting on the $n$ data qubits. These specify the ``inner structure'' of the code that codewords (i.e.\ valid logical quantum states) should abide by: for any stabilizer check operator $K$ in a given stabilizer QEC code, any codeword $\ket{\psi}_L$ should meet the condition $K \ket{\psi}_L = +1 \ket{\psi}_L = (-1)^m \ket{\psi}_L$ with $m=0$. 
Therefore, measuring a stabilizer check operator should always give you an ``$m=0$'' outcome and obtaining an ``$m=1$'' outcome corresponds to an erroneous state.
The bit string of all the stabilizer measurement outcomes is the error syndrome that a decoder uses to (ideally) identify  the most likely error that occurred and correct it.

More general types of codes, such as the subsystem codes~\cite{bacon2006operator}, the Floquet codes and the instantaneous stabilizer group codes~\cite{hastings2021dynamically, townsend2023floquetifying} operate differently but are still based on the measurements of stabilizer check operators. The central QEC properties originate from these stabilizer measurements and being able to decode them (efficiently) to identify errors.
Throughout this paper, we will use the $d \times d$ rotated surface code~\cite{bombin2007optimal}, which is a $\llbracket d^2, 1, d \rrbracket$ QEC code, as an illustrative example for our architecture.

The Tanner graph of a QEC code is a bipartite graph $G_T = (V_T, E_T)$, with $V_T = V_T^{(d)} \sqcup V_T^{(c)}$,
with properties we will now describe.
Vertices in $V_T^{(d)}$ correspond to physical \emph{data} qubits, while vertices in $V_T^{(c)}$, denoted check vertices, correspond to stabilizer check operators (which can be later associated to physical \emph{check} qubits).
The edge set can be decomposed in three disjoint sets $E_{T} = E_{T}^{(X)} \sqcup E_{T}^{(Y)} \sqcup E_{T}^{(Z)}$. An undirected edge $(d_i, c_j) \in E_{T}^{(A)}$ (for $A \in \{X, Y, Z\}$) connects a check vertex $c_j \in V_T^{(c)}$ with a data vertex $d_i \in V_T^{(d)}$ in the Tanner graph whenever the associated stabilizer check operator acts non-trivially on the data qubit through the Pauli operator $A$ on the QEC code. For example, if $\{d_1, d_2, d_3\}$ is the set of neighbors of a check vertex $c_j \in V_T^{(c)}$, with respective edges in $E_{T}^{(X)}, E_{T}^{(Y)}, E_{T}^{(Z)}$, then the associated stabilizer check is $K_{c_j} = X_{d_1} Y_{d_2} Z_{d_3}$.
Figure~\ref{fig_full}(a) represents the Tanner graph of the $d \times d$ rotated surface code~\cite{bombin2007optimal}.

A Calderbank-Shor-Steane (CSS) code has only $X$-type or $Z$-type stabilizer check operators; i.e.\ stabilizer check operators consist either entirely of $X$s and $I$s or of $Z$s and $I$s. Therefore, for any check vertex $c_j\in V_T^{(c)}$, its incident edge set
$E_T(c_j) = \{(d_i, c_j) \in E_{T}\}$
is either included in $E_{T}^{(X)}$ or in $E_{T}^{(Z)}$ ($E_{T}^{(Y)} = \emptyset$). We can thus divide $V_T^{(c)}$ into two disjoint subsets $V_T^{(c, X)}$ and $V_T^{(c, Z)}$ for $X$-type and $Z$-type stabilizer check operators respectively. From $G_T$ we can thus obtain two subgraphs, $G_T^{(X)} = (V_T^{(d)}  \sqcup V_T^{(c, X)}, E_T^{(X)})$ and $G_T^{(Z)} = (V_T^{(d)} \sqcup V_T^{(c, Z)}, E_T^{(Z)})$, also known as (classical) Tanner graphs.
In fact these graphs fully specify a CSS code and can also be used to design efficient decoders~\cite{connolly2022fast, delfosse2022toward}.
The low-density parity check property of a family of code implies that the Tanner graph of each code from this family has a bounded degree, i.e.\ each vertex has a maximum number of neighbors.
For example, the family of $d \times d$ rotated surface codes is not only CSS but also LDPC, since $\forall d \in \mathbb{N}, \forall v \in V_T, |E_T(v)| \leq 4$.

\subsection{Stabilizer measurements with RUS gate}

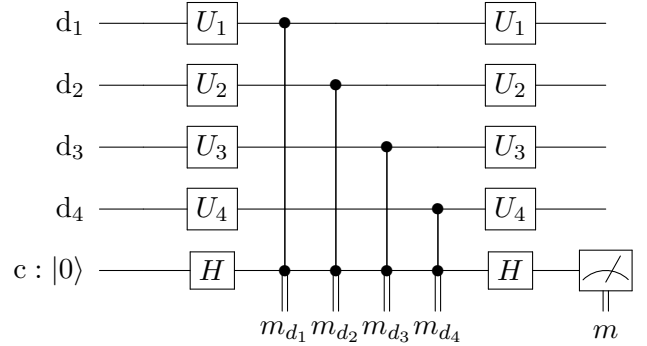
\begin{figure}[thbp]
  \centering
  \[
  \Qcircuit @C=1.5em @R=.7em {
  & & \lstick{\rm d_1} & \qw & \gate{U_{1}} & \ctrl{4} & \qw  & \qw & \qw & \gate{U_{1}} & \qw & \\
  & & \lstick{\rm d_2} & \qw & \gate{U_{2}} & \qw & \ctrl{3} & \qw & \qw & \gate{U_{2}} & \qw  & \\
  & & \lstick{\rm d_3} & \qw & \gate{U_{3}} & \qw & \qw & \ctrl{2} & \qw & \gate{U_{3}} & \qw  & \\
  & & \lstick{\rm d_4} & \qw & \gate{U_{4}} & \qw & \qw & \qw & \ctrl{1} & \gate{U_{4}} & \qw & \\
  & & \lstick{\rm c: \ket{0}} & \qw & \gate{H} & \ctrl{-4} & \ctrl{-3} &  \ctrl{-2} & \ctrl{-1} & \gate{H} & \meter & \\
  & &  &   &   &  \cwx[-1]  & \cwx[-1]   &   \cwx[-1]  &  \cwx[-1] &   &  \cwx[-1] &   \\
  & &  &   &   &  m_{d_1}  & m_{d_2}   &   m_{d_3}  &  m_{d_4} &   &  m &   }
  \]
  \caption{
  General quantum circuit for an indirect stabilizer measurement using a check qubit ``c'' (illustrated for a weight-4 stabilizer check operator). We measure $K_c=A_{d_1} \otimes A_{d_2} \otimes A_{d_3} \otimes A_{d_4}$ by using this circuit and by setting $U_{i}$ to be $H = H_{XZ} = (X + Z)/ \sqrt2$ when $A_{d_i}=X_{d_i}$,  by setting it to $H_{YZ} = (Y + Z)/ \sqrt2$ when $A_{d_i}=Y_{d_i}$, and by setting it to $I_{d_i}$ when $A_{d_i}=Z_{d_i}$. If the measurement outcome is $m=0$ (respectively $m=1$), the system is projected onto the subspace stabilized by $K_c$ (resp. $-K_c$). In this circuit, each conventional CZ gate is replaced by a RUS CZ gate. To indicate a RUS CZ, we decorate a usual CZ with a classical wire that indicates the heralding outcome $m_{d_i}$ which can either herald a success, a failure or an abort signal.
  }
  \label{fig:indirect_meas_rus}
\end{figure}

A crucial part of FTQC revolves around the accurate measurements of stabilizer check operators. 
They are usually challenging to measure directly since they involve multiple qubits. 
As a workaround, stabilizer measurements  are often performed indirectly  by leveraging check qubits. 
The quantum circuit for indirectly measuring a stabilizer check operator $K$ with the assistance of a check qubit ``c'' is illustrated in Fig.~\ref{fig:indirect_meas_rus}.
In this figure, the CZ gate representation is unconventional as they are RUS CZ gates. For the time being, we can set aside this particular detail.
This quantum circuit effectively projects onto the subspace stabilized by $(-1)^m K_c$, where $m \in \{0, 1\}$ is the measurement outcome of the check qubit.
This outcome is subsequently used by the decoder to detect and correct errors in the quantum computation process.

In the context of the SPOQC architecture, deterministic CZ gates are unavailable, and we substitute them with RUS CZ gates. These gates are elaborated upon in Sec.~\ref{sec_rus_gate}.
Despite their inherent probabilistic nature, RUS CZ gates offer the advantage of being \emph{heralded} gates.
This means that we can ascertain whether a gate has succeeded, failed or been aborted, providing valuable information for the decoder. Consequently, a failed or aborted gate can be treated as a heralded error, which typically presents a more manageable decoding problem because the decoder can potentially harness not only the measurement outcome $m$ but also the heralded outcomes of each RUS CZ gate $m_i$ corresponding to the target data qubit involved in the RUS CZ gate with the qubit ``c'' (see Fig.~\ref{fig:indirect_meas_rus}).

\subsection{Modules of the physical architecture}

In the following, 
we present the physical layout of the SPOQC architecture, which is designed with a modular structure.
We provide a description of each of its primary modules.

\paragraph{Quantum emitter module.}
A quantum emitter module includes a quantum emitter with a spin qubit, and all its control components for single-photon emission and spin control.
We will assume that we have full $SU(2)$ control on the spin qubit and that it can emit spin-entangled photons through the emission gate $E_{\rm qe, ph} = \ket{0_{\rm qe},0_{\rm ph}}\bra{0_{\rm qe}} + \ket{1_{\rm qe},1_{\rm ph}}\bra{1_{\rm qe}}$, or equivalent~\cite{thomas2022efficient, cogan2023deterministic, coste2023high, appel2022entangling, meng2023deterministic}. The quantum emitter sends photons in a privileged single direction, optimizing the efficient collection of photonic qubits (for example using a fiber)~\cite{thomas2021bright, tomm2021bright, uppu2020scalable, bhaskar2020experimental, thomas2022efficient}.
The control components within the system are operated through classical information processing. For instance, we rely on classical information to determine whether to proceed with emitting a new spin-entangled photon and initiating a new trial. This decision is contingent upon receiving information that heralds either the success, repeat, or failure outcome of a RUS gate trial.

\paragraph{Quantum channel.}
The propagating photons are transferred from one module to another thanks to quantum channels that can be implemented using optical fibers or free-space components.

\paragraph{Photon router module.}
The photon router module is an active component that routes photons from a specified input to a designated output~\cite{lenzini2017active, pont2022quantifying, hansen2023single}. In our modules, we employ a $1 \times N$ router configuration, where photons originating from a single mode can be routed to any of the $N$ output modes.
$N$ is related to the maximum degree of the Tanner graph, which is bounded, by design, with LDPC codes.
The routing strategy depends on information transmitted to the router via a classical channel.

\paragraph{RUS gate module.}

A RUS gate module takes as input two photonic qubits and performs a linear-optical transformation followed by photon detection. It sends the detector measurement outcome through classical channels.

\subsection{Macroscopic layout of the physical architecture}

In the following, we explore how we can exploit the fault-tolerant properties of a given QEC code using an arrangement of the preceding modules.

Based on the QEC Tanner graph and the previously introduced modules, we propose the following layout. 
The data and check qubits of a QEC code are encoded using the spin qubits of quantum emitters. 
Therefore, for each data or check qubit, corresponding to a vertex in the Tanner graph, we use a single quantum emitter module. 
Each quantum emitter module is connected to a photon router module thanks to a quantum channel (e.g. a fiber).

The SPOQC architecture is suited for LDPC codes because the largest number of photon router outputs necessary to perform every stabilizer measurement circuit corresponds to the Tanner graph's maximum degree.
Yet, increasing the number of outputs of each photon router could also offer other advantages such as enabling fault-tolerant gates \cite{moussa2016transversal}, single-shot error correction~\cite{campbell2019theory}, or increasing the compilation speed~\cite{litinski2022active}. 

If there is an edge between vertices $c_j\in V_T^{(c)}$ and $d_i \in V_T^{(d)}$, 
a RUS gate module should be connected to one of the output ports of the photon router modules of both the corresponding check qubit ${\rm c}_j$ and data qubit ${\rm d}_i$. 
Note that the edge type, $X$, $Y$, $Z$, can be decided by applying a Hadamard gate $H = H_{XZ} =  (X + Z) / \sqrt2$, a Y-Hadamard gate $H_{YZ} = (Y + Z) / \sqrt2$ or the identity gate on qubit ${\rm d}_i$ just before and immediately after the RUS CZ gate as shown in Fig.~\ref{fig:indirect_meas_rus}.

The example given in Fig.~\ref{fig_full}(b) shows this architecture implemented for a local code – the $3 \times 3$ rotated surface code whose Tanner graph was previously introduced in Fig.~\ref{fig_full}(a). Note, however, that contrary to other architectures, the SPOQC architecture uses long-range optical links and is thus not restricted to codes respecting locality constraints. This allows it to implement any QEC code, and in particular 
any LDPC code.

\section{Repeat-until-success linear-optical gates}
\label{sec_rus_gate}

\subsection{General overview}

In this section, we focus on the repeat-until-success (RUS) gates which are central to this architecture.
In its broadest sense~\cite{paetznick2013repeat, shah2013ancilla},
a RUS implementation of a quantum gate $U$ relies on the repetition of a non-deterministic quantum circuit $C$ involving measurements. $C$ should be repeated until one of the measurement outcomes corresponding to the successful implementation of $U$ is obtained. Critically, the other measurement outcomes should, up to some local correction, correspond to the identity gate so that it can be repeated indefinitely.

Here, however, following the original terminology~\cite{bose1999proposal,lim2005repeat}, the term RUS gate will refer specifically to the CZ gate, whose non-deterministic circuit is realized by the emission of entangled photons passing through a linear optical circuit before being detected.

Note that linear-optical RUS gates typically have higher success rates for a given amount of loss than ancilla-assisted fusion gates and boosted fusion gates~\cite{hilaire2022near} when compared to other common linear-optical gates used for generating entanglement. This is why we can expect an improved loss tolerance compared to other implementations without sacrificing the architecture's simplicity.
RUS gates have been proposed to generate graph states~\cite{lim2006repeat}, including the ``Raussendorf-Harrington-Goyal'' lattice~\cite{raussendorf2007topological, li2010fault}, where the qubits in the graph states are the quantum-emitter qubits. A recent protocol~\cite{bell2022protocol} uses a linear-optical RUS gate to generate a $\ket{W}$ entangled state on $d$ quantum emitters which can then be used to generate photonic qudit GHZ states near-deterministically.

Below, we provide an in-depth presentation of these RUS gates.

\subsection{Quantum emitter}

In addition, we consider quantum emitters that have a degree of freedom, such as a spin on which we can encode a qubit, $\ket{0}_{\rm qe} = \ket{\uparrow}_{\rm qe}$, $\ket{1}_{\rm qe} = \ket{\downarrow}_{\rm qe}$. 
Hereafter, note that while we informally refer to this degree of freedom as a ``spin'', it can represent any pair of energy-level states within the quantum emitter.
We also consider that one of the single-photon degrees of freedom, such as its polarization or its path, encodes a qubit.  
This degree of freedom can also be entangled with the spin state upon emission, with an emission process described by the operator
$$
E_{\rm qe, ph} = \ket{0}_{\rm qe} \ket{0}_{\rm ph} \bra{0}_{\rm qe} + \ket{1}_{\rm qe} \ket{1}_{\rm ph} \bra{1}_{\rm qe}.
$$
This operator permits the emission of spin-entangled photons~\cite{lindner2009proposal}. 
For example, if the spin is initialized to $\ket{+}_{\rm qe} = \ket{0}_{\rm qe} + \ket{1}_{\rm qe}$ (omitting state normalization), after the emission of a photon, the resulting state is $E_{\rm qe, ph}\ket{+}_{\rm qe} = \ket{0}_{\rm qe} \ket{0}_{\rm ph} + \ket{1}_{\rm qe} \ket{1}_{\rm ph}$. 

\begin{figure}[thbp]
    \centering
    \includegraphics[width=\columnwidth]{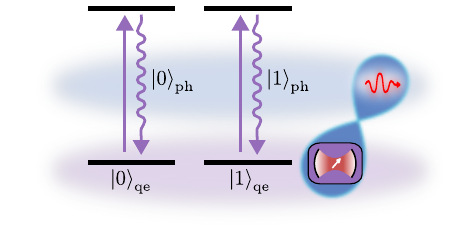}
    \caption{
    Quantum emitter level structure. The optical transitions, represented in violet, enable the emission of a spin-entangled photon following operator $E_{\rm qe,ph}$.}
    \label{fig_quantum_emitter}
\end{figure}

A quantum emitter that naturally allows the emission of spin-entangled photons through the operator $E_{\rm qe, ph}$ is a singly-charged quantum dot whose energy levels form a four-energy-level system with two optical transitions as shown in Fig.~\ref{fig_quantum_emitter}~\cite{schwartz2016deterministic, cogan2023deterministic, coste2023high}. In this system, the photonic qubits are encoded using the polarization degree of freedom.
However, this emission process is by no means restricted only to such a level structure and can be adapted to many other configurations and most quantum emitter platforms, such as other types of quantum dots, atoms, superconducting circuits,  and color defects~\cite{thomas2022efficient, appel2022entangling, lee2019quantum, vezvaee2022deterministic, tiurev2022high, vasconcelos2020scalable, besse2020realizing,liu2022proposal}.

\subsection{Photonic qubits}
The quantum emitter described in Fig.~\ref{fig_quantum_emitter} emits photons entangled in polarization but it is straightforward to convert polarization-encoded photonic qubits into path-encoded photonic qubits, for example using a polarizing beamsplitter.
In path encoding,
$\ket{0}_{\rm ph} = \ket{1,0}\rangle_{\rm ph}$, $\ket{1}_{\rm ph} = \ket{0,1}\rangle_{\rm ph}$, where $\ket{i,j}\rangle$ corresponds to having $i$ ($j$) photons in the first (second) spatial mode. We will use the path encoding for the linear-optical interferometer description in the following.

\subsection{Linear-optical interferometer}

The linear-optical interferometer presented in Fig.~\ref{fig_full}(c) implements the following unitary transformation on the photonic modes (a variant of that found in Ref.~\cite{lim2005repeat})

\begin{equation}\label{eq_unitary_rus}
    U = \frac{1}{2}
    \begin{pmatrix}
        1 & 1 & 1 & 1\\
        1 & 1 & -1 & -1\\
        1 & -1 & -i &  i\\
        1 & -1 & i & -i
    \end{pmatrix}.     
\end{equation}

which we use to perform a (CZ) RUS gate.
This interferometer can be seen as a fusion gate~\cite{bartolucci2023fusion}, as its goal is to perform a joint measurement on the dual-rail-encoded photons. However, here, we are more interested in its indirect effect on the spin state of the emitters that produced the two photons.

\noindent The input modes $0$ and $1$ (respectively $2$, $3$) correspond to the dual-rail modes of the photonic qubits emitted by quantum emitter $a$ (respectively quantum emitter $b$).
Four photon-number-resolving (PNR) detectors are positioned at the output modes of the interferometers. 
We denote by $(k,l)$ a detection pattern where detectors in output modes $k$ and $l$ each detect a single photon. Note that $(k, l)$ is also equivalent to the detection pattern $(l, k)$. A detection event $(k,k)$ corresponds to the detector in output mode $k$ detecting two photons.

We provide a comprehensive description of the functioning of the RUS gate in the following and more detailed analysis from the stabilizer formalism point of view in Appendix \ref{RUS_channels}.

\paragraph{Emission step.}
In each trial, every quantum emitter sends an entangled photon through the operation $E_{\rm qe, ph}$ into the linear-optical interferometer. The photons are detected at the output modes of the interferometers. The spin gate that has been performed depends on the detection pattern. 

\paragraph{Successful measurement outcomes.} Some measurement outcomes such as $(0,2)$, i.e.\ a photon detected 
at output mode $0$ and a photon detected 
at output mode $2$, correspond to a spin CZ gate up to single-qubit unitary corrections, e.g. $S_a {S_b}^\dag$ in the $(0,2)$ case. Here, $S_i = \sqrt{Z_i}$ is the phase gate on qubit $i$. When performing these corrections, we obtain a successful spin CZ gate and the RUS gate has succeeded at this trial.

\paragraph{``Repeat'' measurement outcomes.} 
Other measurement outcomes such as $(2, 2)$, i.e.\ two photons detected at output mode $2$, correspond to an identity gate up to correction, e.g. $Z_a Z_b$ for the $(2,2)$ case. In that case, we have not performed the desired CZ gate but we can try again by re-emitting photons into the interferometer until we obtain a successful measurement outcome.

\begin{table*}[!ht]
    \centering
    \begin{tabular}{|c||c|c|c|c||c|c|c|c|c|c||c|}
        \hline
        Detection & $(0,2)$ & $(1,3)$ & $(0,3)$ & $(1, 2)$ & $(0,0)$ & $(1,1)$ & $(0,1)$ & $(2,2)$ & $(3,3)$ & $(2, 3)$ & $\mathcal F$  \\
        \hline
        Probability & $\frac{\eta_a \eta_b}{8}$ & $\frac{\eta_a \eta_b}{8}$ & $\frac{\eta_a \eta_b}{8}$ & $\frac{\eta_a \eta_b}{8}$ & $\frac{\eta_a \eta_b}{8}$ & $\frac{\eta_a \eta_b}{8}$ & 0 & $\frac{\eta_a \eta_b}{8}$ & $\frac{\eta_a \eta_b}{8}$ & 0 & $1 - \eta_a \eta_b$  \\
        \hline
        Corrections & \multicolumn{2}{c|}{$S_a {S_b}^\dag$} & \multicolumn{2}{c||}{${S_a}^\dag {S_b}$} & \multicolumn{3}{c|}{Id} & \multicolumn{3}{c||}{$Z_a Z_b$} & Id  \\
        \hline
        Spin Gate & \multicolumn{4}{c||}{$CZ_{a,b}$} & \multicolumn{6}{c||}{Id} & $C_{RUS, f}$  \\
        \hline
    \end{tabular}
    \caption{RUS gate detection patterns and associated transformations.
    The first row corresponds to the detection pattern that has been observed, noting that the last column $\mathcal F$ corresponds to all cases for which one or both photons have been lost. The second row is the detection event probability (for mixed input state and indistinguishable photons). The third and fourth rows correspond to the spin corrections and the gates between spins that are performed.}
    \label{tab:RUS_sumup}
\end{table*}

\paragraph{Gate failure.}
A RUS gate fails at a given trial, denoted as an $\mathcal F$ detection pattern, if strictly less than two photons are detected.
Such an event can only happen in the presence of photon loss (including imperfect efficiency of the detectors). In that case, we lose partial information of the spin states, and the corresponding channel corresponds to a spin phase erasure applied to the two quantum emitters,
$$C_{RUS,f} = C_{Z_a} C_{Z_b} = C_{Z_b} C_{Z_a}$$ 
where
$$C_{Z_i}(\rho) \vcentcolon= \frac{1}{2} \left( \rho + Z_i \rho  Z_i\right).$$

Indeed, it is easy to show that $C_{Z_i}(\rho) = {\rm Tr}_{\rm ph}(E_{i, {\rm ph}}\rho {E_{i, {\rm ph}}}^\dag)$, i.e.\ emitting a photon from quantum emitter $i$ and tracing this photon out to denote it being lost. If we lose both photons it is clear that we apply $C_{RUS,f}$. If we lose only one photon, the linear-optical interferometer erases the ``which-path'' information of the detected photon, so that we don't know which quantum emitter produced it. 
Therefore, in that case as well, we should apply $C_{RUS,f}$.

Detection events along with their resulting correction unitaries and spin gates  are summarized in Table \ref{tab:RUS_sumup}~\footnote{In principle, for perfectly indistinguishable photons, it is impossible to obtain a $(0, 1)$ or a $(2,3)$ detection outcome. However, this is possible for distinguishable sources, and they correspond to an identity gate (up to a $Z_a Z_b $ outcome for the $(2, 3)$ detection).}.
Each trial of the RUS gate can either succeed, fail, or reach a repeat outcome. The success and failure outcomes terminate the execution of the RUS gate at this trial, but the repeat case calls for another trial. Since a realistic RUS gate cannot take infinite time, we should set a maximum number of trials $k$. If we obtain a full sequence of $k$ repeat outcomes, we should abort and not complete the CZ gate. In that case, an identity gate is performed instead.\footnote{The terminology used here differs from the one that is usual for fusion gates. A RUS gate ``failure'' corresponds to a ``photon loss'' case in fusion gate terminology. A RUS ``repeat'' case corresponds to a fusion gate ``failure'' case, but can be tried again (which is usually not the case for fusion gates). If all the RUS trials yield a RUS ``repeat'' outcome, we have an ``abort'' case.}

\subsection{Success rate}

From Table \ref{tab:RUS_sumup}, it is clear that each gate trial has a success rate of $p_s = \eta_a \eta_b / 2$, a ``repeat'' rate of $p_r = \eta_a \eta_b/2$ and a failure probability of $p_f = 1 - \eta_a \eta_b$, where $\eta_i$ is the end-to-end transmission efficiency (i.e.\ from photon emission to detection) of a photon emitted by quantum emitter $i$. Whenever a trial yields a repeat pattern, we can try again. Therefore the overall RUS gate success rate is given by:
\begin{equation}
    \begin{aligned}
        P_{\rm RUS,s}(\eta_a, \eta_b, k) & = p_s \sum_{n=0}^{k-1} (p_r)^n \\ 
        &= \frac{\eta_a \eta_b}{2} \frac{1 - (\eta_a \eta_b/2)^k}{1 - \eta_a \eta_b/2} \\ 
        & \xrightarrow[k \to +\infty]{}\frac{\eta_a \eta_b}{2 - \eta_a \eta_b}    
    \end{aligned}
\end{equation}
where $k$ is the maximum number of trials allowed, since in practice a gate cannot take infinite time.
Similarly, the failure rate is
\begin{equation}
    \begin{aligned}
        P_{\rm RUS,f}(\eta_a, \eta_b, k) & = p_f \sum_{n=0}^{k-1} (p_r)^n \\
        &= (1 - \eta_a \eta_b) \frac{1 - (\eta_a \eta_b/2)^k}{1 - \eta_a \eta_b/2} \\ 
        & \xrightarrow[k \to +\infty]{}\frac{2- 2\eta_a \eta_b}{2 - \eta_a \eta_b},
    \end{aligned}
\end{equation}
and the aborted rate (where no gate is applied) is
\begin{equation}
    \begin{aligned}
        P_{\rm RUS,a}(\eta_a, \eta_b, k) & = 1 - \left[ P_{\rm RUS,s}+ P_{\rm RUS,f} \right](\eta_a, \eta_b, k) \\
        & \xrightarrow[k \to +\infty]{} 0.
    \end{aligned}
\end{equation}

Since we maximize the gate success probability by repeating it until it succeeds or fails, the RUS gates are quite efficient at performing a gate between remote qubits using linear optics. However, between each trial step of the gate, we need a round of communication between the quantum emitter modules and the RUS module. Indeed, a pair of photons should be transferred from a quantum emitter towards the detectors and the classical information of the measurement outcome should be transferred back to the quantum emitter's nodes to decide whether we should or not proceed with a next trial. Given a distance $L_0$ between the quantum emitter and the detectors, this implies a waiting time of at least $2 L_0 / c$, with $c$ the speed of light, between two trials. Therefore, to avoid infinite gate time which are not feasible in practice, we should allocate a maximum number $k$ of trials. In that case, if the gate  is limited by the communication time delays, it lasts at least $2k L_0/c$.

\section{Numerical results} \label{sec_results}

In this section, we present the numerical results that we obtained with the SPOQC architecture. While this architecture can operate for any QEC stabilizer code, we consider rotated surface codes with a Minimum Weight Perfect Matching (MWPM) decoder to facilitate the comparison with other existing architectures. We use Stim~\cite{gidney2021stim} and Pymatching~\cite{higgott2022pymatching, higgott2023sparse} for our simulations.

We center our analysis on the error thresholds and we estimate them by calculating the crossing point of logical error curves for different code distances~\cite{stace2010error, paesani2022high}. 
Because we use distances up to 13, the values provided for thresholds are lower bounds on the actual thresholds.
Our analysis centers on a physically-motivated error model that focuses on photonic errors and spin decoherence. 
In particular, we focus on the loss threshold and the spin decoherence time to reveal an interesting interplay between these two types of errors that arises from the physical implementation of an FT scheme with probabilistic RUS gates.
We also consider partial distinguishability between the photons which causes errors in the RUS gates. To the best of our knowledge, this is the first time that an FT threshold has been derived for such an intrinsically photonic error.
This model allows us to gain a comprehensive understanding of how these specific errors affect the performance of our architecture.
Details about these error simulations can be found in Appendix~\ref{sec_physical_noise}.

\subsection{Independent RUS gate thresholds}

\begin{figure}
    \centering

    \hspace{-8.3cm} a)
    
    \includegraphics[width=\columnwidth]{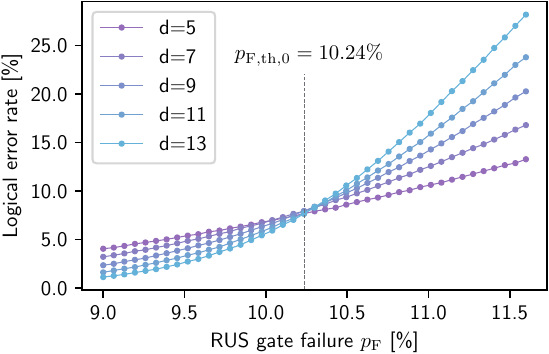}
    
    \vspace{1cm}
    \hspace{-8.3cm} b)
    
    \includegraphics[width=\columnwidth]{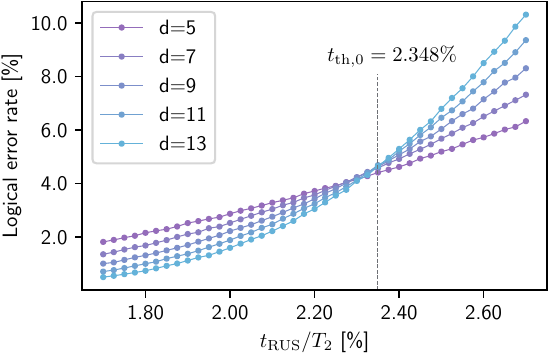}

    \vspace{1cm}
    \hspace{-8.3cm} c)

    \includegraphics[width=\columnwidth]{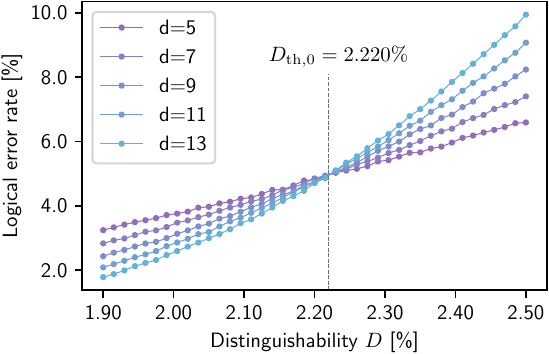}
    \caption{
    Logical error rate for different distance values vs (a) the RUS gate failure probability $p_F$, (b) the decoherence time of the spin $t_{\rm RUS} / T_2$, and (c) distinguishability of photons.
    }
    \label{fig_t_threshold}
\end{figure}

\paragraph{RUS gate failure threshold.}
We first start by evaluating the threshold in terms of RUS gate failure or abort probability $p_{F}(\eta_a, \eta_b, k) = 1 - P_{\rm RUS,s}(\eta_a, \eta_b, k)$. 
Abort cases for RUS gates arise when the gate was neither a success nor a failure while the maximum number of allowed trials was exhausted resulting in an identity gate applied on the spins. 
For simplicity when running the simulations, we treat abort cases the same way as failure cases and apply the error $C_{\rm RUS,f}(\rho)$, which thus overestimates the amount of noise. Given that a realistic RUS gate depends on the maximum allowed number of trials $k$, this allows it to be evaluated simultaneously for all $k$. Assuming no spin decoherence $T_2 = + \infty$, and perfectly indistinguishable photons, we find the gate failure threshold $p_{\rm F, th, 0} = 10.24 \pm 0.04\%$ in Fig. \ref{fig_t_threshold} (a) for an implementation of the SPOQC architecture with a rotated surface code. The interested reader can find details about our simulation methods in Appendix~\ref{sec_notes_simu}. Assuming infinite decoherence time and uniform loss, we can in principle take as many trials as needed $k \to \infty$ and obtain a single-photon loss threshold approaching $1-\eta_{\rm th} = 2.75 \pm 0.02\%$.

\paragraph{Spin decoherence time.}
A spin qubit is usually described by two figures of merit, its relaxation time $T_1$ and its decoherence time $T_2$.
In the following, we consider that spin decoherence is the main source of errors of our quantum emitters: $T_2 \ll T_1$, which is usually the case for quantum emitter platforms. We will not consider the spin relaxation time as a source of error ($T_1 = +\infty$) and only focus on the decoherence time $T_2$.
It is meaningful to express this decoherence time relative to the time required to perform a RUS gate  $t_{\rm RUS}$.
We should thus obtain a threshold for the ratio $t_{\rm RUS} / T_2$: below a threshold value $t_{\rm th}$, we are in the FT regime. 
Note that here again, we consider the time $t_{\rm RUS}$ for a full RUS gate so that the threshold is to be divided by the number of trials $k$ of the RUS gate, as $t_{\rm RUS} = k t_{\rm trial}$.
As shown in Fig.~\ref{fig_t_threshold} (b), assuming no photon losses ($p_{\rm F} = 0$) and perfectly indistinguishable photons, we find the FT condition $t_{\rm RUS} / T_2 < t_{\rm th, 0} = 2.348 \pm 0.009 \%$.

\paragraph{Distinguishable photons.}
Linear-optical entangling gates, such as RUS gates, heavily rely on having indistinguishable photonic qubits, i.e. photons that are indistinguishable in every degree of freedom other than the one used for their qubit encoding. The presence of partially distinguishable photons introduces errors while performing the RUS gate -- errors which could obviously hinder the computation.
We let $D$ denote the distinguishability between photons, assuming it to be the same for all photons from all emitters.
The error channel caused by $D$ is approximated by 
$$
C_{D, err}(\rho) \vcentcolon= (1-D) \rho + D C_{Z_a} C_{Z_b} (\rho)
$$ 
and we refer interested readers to Appendix.~\ref{subsec_distinguishability} for a detailed discussion on this result. Assuming no spin errors and no photon loss, we find that the SPOQC architecture can tolerate up to $D_{\rm th, 0} = 2.220 \pm 0.013 \%$ of distinguishable photons, see Fig.~\ref{fig_t_threshold} (c).

\subsection{Multi-error RUS thresholds}

\begin{figure}[thbp]
    \centering
    \includegraphics[width=\columnwidth,trim={2.94cm 0 1cm 0}]{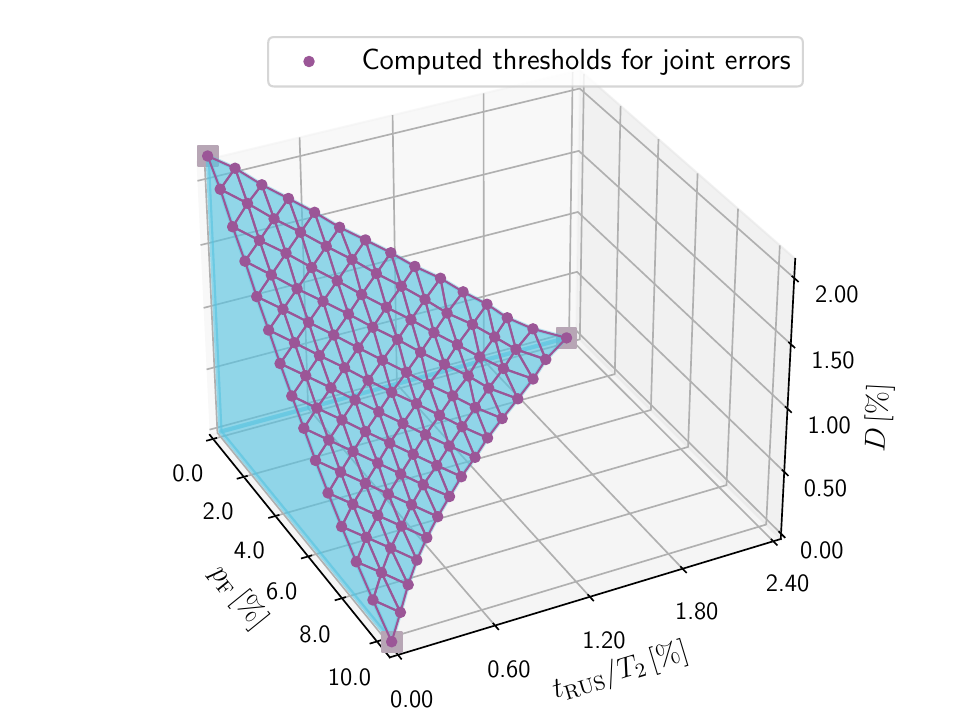}
    \caption{Fault-tolerant region for the SPOQC architecture in terms of $t_{\rm RUS} / T_2$, RUS gate failure $p_F$, and distinguishability $D$. The points highlighted with the light violet squares correspond to the individual error thresholds computed previously.}
    \label{fig_RUS_threshold}
\end{figure}

In the previous section, we have assumed independent errors: either solely loss, photon distinguishability, or decoherence. In practice, all errors can happen simultaneously and, 
intuitively, the more we have of one type of error the less we can correct of the other types. Given a vector of errors $\vec m = (t = t_{\rm RUS} / T_2, p_F, D)$, we want to know whether it lies in the correctable region or not.

Through a numerical exploration detailed in Appendix~\ref{sec_notes_simu}, we obtain the curve plotted in Fig.~\ref{fig_RUS_threshold}, which exhibits the correctable region (in light blue) for RUS gate failure, distinguishable photons, and spin decoherence time.

\subsection{Physically relevant error threshold}

\subsubsection{With standard RUS gates}

\begin{figure}[thbp]
    \centering
    \includegraphics[width=\columnwidth]{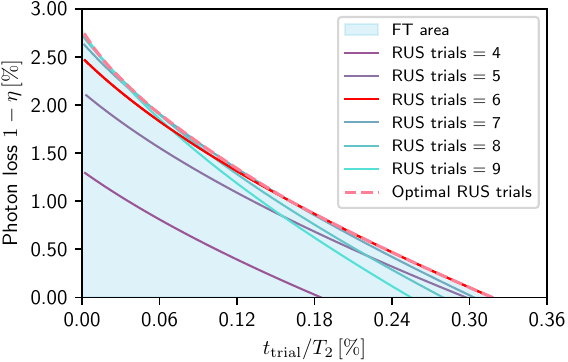}
    \caption{Fault-tolerant region of the SPOQC architecture in terms of $t_{\rm trial} / T_2$ and single-photon loss $\varepsilon$ as a function of the number $k$ of trial of the RUS gates. The optimal curve corresponds to the maximum fault-tolerance optimized over $k$. The curve $k=6$ is highlighted in red because it demonstrates remarkable balance between photon loss and decoherence time of the spin.}
    \label{fig_physical_threshold}
\end{figure}

So far, we have exhibited the correctable region parameterized by the RUS gate parameters. Note that taking a larger number of trials $k$ would result in a better single-photon loss tolerance. However, increasing $k$ would also mean having significantly more stringent decoherence time requirements since the RUS gate total time would be larger. We aim at quantifying this interplay for the RUS gates as described in Section~\ref{sec_rus_gate}.

To do so, we interpolate the border of the correctable area in Fig.~\ref{fig_RUS_threshold} for a distinguishability $D=0$.
We use this curve to estimate, given a value of $k$ and a single-photon loss $1 -\eta$, what is the maximum ratio $t_{\rm th} =t_{\rm RUS} /T_2$ tolerated by the architecture.
From that, given a single-photon loss probability, we obtain straightforwardly the maximum trial time $t_{\rm trial} = t_{\rm th} / k$ allowed in order to remain in the FT regime. Repeating this for many $k$ values results in finding the optimal FT region of loss and spin decoherence errors. The results are displayed in Fig.~\ref{fig_physical_threshold}.

A first important conclusion that we can draw from Fig.~\ref{fig_physical_threshold} is that our scheme cannot be FT for values of $k$ below 3. Indeed, even without loss the RUS gate fails or aborts with at least a probability $2^{-k}$, and for $k \leq 3$, $p_{\rm F, th, 0} \leq 2^{-k}$. Therefore, it is impossible to reach the maximum RUS gate failure threshold if we do not allocate at least 4 trials per RUS gate.

A second important remark is that  the optimal number of trials $k$ depends on both the errors and their nature. Below the FT thresholds, systems more sensitive to photon loss would typically require a larger number of trials $k$ for their RUS gates than systems more sensitive to spin decoherence. Also, even for small $t_{\rm trial} / T_2$, having $k \geq 8$ only marginally increases the loss tolerance, and the correctable area for $k=6$ covers all of the correctable areas for smaller values of $k$ and almost all the global FT region (a point is in the globally FT region if and only if for some $k$ it is correctable) delimited by the optimal RUS trial curve. 

We obtain that the SPOQC architecture can tolerate physical errors included in the FT area of Fig~\ref{fig_physical_threshold}. With the rotated surface code, the SPOQC architecture  tolerates up to $2.75\pm0.02 \%$ photon loss and requires that a RUS gate trial time takes at most $0.318 \%$ of the decoherence time $T_2$. 
However, it should be noted that the more photon loss the more trials are required to reach the loss threshold and thus the shorter we require each trial RUS gate to take compared to the spin decoherence time.

\subsubsection{With hybrid RUS gates}

As we emphasized by Fig.~\ref{fig_physical_threshold}, RUS gates offer the best robustness against photon loss.
However implementing a RUS gate can take a relatively long time as it requires sending photons one-by-one so the duration of each trial is the time of photon emission, transfer, and detection with additional time for classical processing and feedforward delays.  
This places stringent requirements on the spin coherence time.
We investigate a slight relaxation of RUS gates where we allow the emitter to send multiple photons in a single trial. 
These photons will be processed by an adaptive linear-optical interferometer. This is detailed in Appendix~\ref{app_hybrid_RUS}.
We call these gates \emph{hybrid} RUS gates. 
They are parameterized by the number of maximum trials $k$ allowed and the number of photons $n$ sent in each trial. 

\begin{figure}[thbp]
    \centering
    \includegraphics[width=\columnwidth]{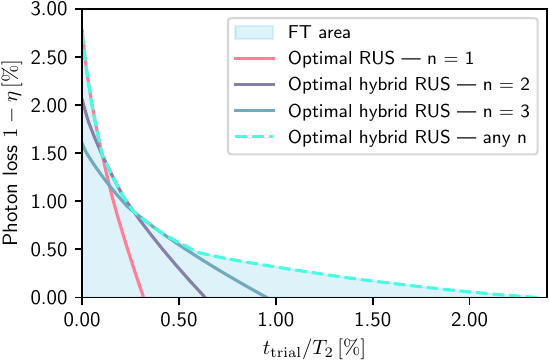}
    \caption{Fault-tolerant area of the SPOQC architecture in terms of $t_{\rm trial} / T_2$ and single-photon loss $\varepsilon$ for boosted fusion gates with varying number $n$ of emitted photons $n$ per trial. These curves have been optimised in the number of trials $k$.}
    \label{fig_boosted_fusion}
\end{figure}

Again we investigate the interplay between photon loss and decoherence time of the spin, but this time for hybrid RUS gates.
Results are shown in Figure~\ref{fig_boosted_fusion}. As expected we observe that at the expense of loss tolerance, we can drastically increase robustness to spin decoherence.

Note that the limit case for hybrid RUS gates in which we allow any number of photons and only one trial corresponds to boosted fusion gates~\cite{hilaire2023linear} and that hybrid RUS gates with only one photon and many trials correspond to the standard RUS gates we introduced earlier.

\section{Comparison with existing architectures}\label{sec_comp}
Regarding comparison with other architectures, we should emphasize that a quantum emitter can have two roles, either as a carrier of quantum information or as a photon source.

Having a non-deterministic two-qubit gate is obviously a limitation compared to platforms with a full set of deterministic gates. Yet, using quantum emitters with long-range optical connections enables the use of non-local links which could help in reducing the FTQC footprint by using high-encoding-rate LDPC codes~\cite{roffe2020decoding, panteleev2022asymptotically, leverrier2022quantum, tillich2013quantum,bravyi2023high}. Such codes require drastically less physical qubits per logical qubit compared to platforms with only local interaction.
Indeed, non-local links between qubits allow for the use of \emph{any} quantum LDPC code that could be appealing to implement, beyond just 2D or 3D codes.
For example an impressive encoding rate of $k/n \to 13/72 \approx 0.18$, i.e.\ approximately 1 logical qubit for every 11 physical qubits (check qubits included) is known to be achievable~\cite{breuckmann2021single}.

Moreover, this architecture could significantly facilitate the error correction process by exploiting single-shot error correction~\cite{campbell2019theory} and speed up the compilation process by implementing active volume strategies that rely on long-range interactions~\cite{litinski2022active}. Furthermore, an architecture based on distant qubits could ensure a better isolation of each qubit, avoiding correlated noise, which is particularly detrimental for error correction~\cite{xu2022distributed, ang2022architectures}.

Recently, Ref.~\cite{lobl2023loss} proposed an architecture to produce a percolated graph state from quantum emitters using photonic probabilistic fusion gates and then renormalizing this graph into an interesting lattice. This scheme leads to a remarkable loss-tolerance threshold ($\approx 6\%$) and could also be envisioned using RUS gates. However, it also comes with the drawback of necessitating a renormalization step, which requires significant resource overhead and more stringent demands on the spin qubit coherence, control, and measurement performances. Indeed, errors in any quantum emitter measured out during the renormalization step may propagate to the data qubits and should thus increase the error rate that needs to be handled using quantum error correction. 

Quantum emitters can also be embedded in photonic architectures, for example as a resource state generator in the fusion-based scheme of Ref.~\cite{bartolucci2023fusion} and related all-photonic schemes~\cite{omkar2022all, pankovich2023high}. We reach similar loss tolerance ($1-\eta_{\rm th,0}\approx 2.75 \%$) to Ref.~\cite{bartolucci2023fusion} when it uses the Shor(2,2) 6-ring resource state and much better performances compared to when it uses the non-encoded 4-star and 6-ring resource states.
Moreover, while we achieve similar loss tolerance, the physical architecture is much simpler since the fusion-based architecture would require highly-reliable generation of Shor(2,2) 6-ring resource states, i.e.\ a 24-qubit photonic resource state with intrinsically small successful generation probability.
Building complex resource states such as these cannot be realized deterministically with non-interacting quantum emitters~\cite{li2022photonic}. To circumvent this issue, a resource state generator should heavily rely on multiplexing to ensure that a resource state is produced with arbitrarily high success probability.  
Such a resource state generator, even one based on quantum-emitter sources of entangled photons, could be quite resource-intensive and would likely require complex optical setups. In comparison, the SPOQC architecture does not rely on multiplexing and thus requires comparably much simpler optical setups.

The original proposal for quantum computing using RUS gates~\cite{lim2005repeat, lim2006repeat} also considered the generation of spin-based graph states, which can be used for FTQC. For example, Ref.~\cite{li2010fault} considered using probabilistic gates to build edges with high probability, resulting in a fault-tolerant 3D lattice.
In that work, each vertex of the graph should correspond to a spin qubit. However, increasing its efficiency requires building a graph state of many spin qubits per vertex. In this respect, the proposal resembles the fusion-based architecture except that the resource state generator should prepare graph states of spin qubits instead of photons, with high success probability.
Such an architecture leads to a much larger resource overhead since it requires multiplexing and more complex spin and optical setups compared to the relative simplicity of the SPOQC architecture.

\section{Discussions}
\label{sec_discussion}
We have presented a modular fault-tolerant quantum computing architecture for quantum-emitter platforms based on photonic repeat-until-success gates. Contrary to most platforms with only nearest neighbor qubit links, here we can have arbitrary physical qubit connectivity since the two-qubit gates are made using photonic links. Therefore, this architecture is compatible with any QEC stabilizer code as it is organized around its Tanner graph, whose connectivity can be arranged with photonics. 
Furthermore, this architecture is compatible with the important class of LDPC codes.

Although we have presented the SPOQC architecture with quantum-dot emitters in mind~\cite{maring2023general}, it can be made compatible with many other physical platforms \cite{atature2018material}. In addition to atomic \cite{thomas2022efficient} and trapped-ion \cite{blinov2004observation, bock2018high} systems, this includes platforms based on proven spin-photon interfaces such as nitrogen-vacancy centers in diamond~\cite{tchebotareva2019entanglement, vasconcelos2020scalable} and emerging emitters like rare earth ions~\cite{siyushev2014coherent, ruskuc2022nuclear} and silicon defects~\cite{higginbottom2022optical, lee2023high}. The SPOQC architecture may also be immediately applicable to many optically-active defects yet to be discovered in emerging materials like hexagonal boron-nitride \cite{castelletto2020hexagonal,kubanek2022coherent, patel2023dynamical}. Adapting SPOQC to the level structure of some of these emitters may require using a time-bin photonic qubit encoding~\cite{tchebotareva2019entanglement}.
In future work, we will explore directions for further enhancing the performance of the architecture. Adaptive strategies can improve the loss tolerance of architectures based on linear-optical gates~\cite{auger2018fault, bell2022optimising,hilaire2023linear, bombin2023increasing}. 
This has recently drastically improved the tolerance to photon loss probability of the fusion-based architecture of Ref.~\cite{bartolucci2023fusion} from 2.7\% to 7.5 \%. 
We can expect to further improve the loss tolerance performance of the SPOQC architecture using similar strategies, for example by deciding whether or not to perform a probabilistic gate based on previous measurement outcomes. 
Moreover, in the current proposal, each RUS gate is performed sequentially and each gate takes a fixed number $k$ of trials. However, these gates do not need to be performed sequentially: we do not need to wait to fully complete a first RUS gate between qubits $a$ and $b$ before starting making one between $a$ and $c$. Instead, we can perform them ``in parallel''.
Besides, instead of having a fixed number of trials per RUS gate, we can, in principle, dynamically allocate the trials to multiple gates and we can envision requiring fewer trials for similar performances: if a RUS gate succeeds, we can dynamically allocate its remaining trials to another gate. We expect this strategy to reduce the time it takes to perform a full error correction cycle and thus be less sensitive to spin decoherence. These are three potential routes for further investigations and improvements of this architecture.

\begin{acknowledgments}
We thank Boris Bourdoncle for fruitful discussions throughout the entire project. We thank Sharon David and Théo Dessertaine, who helped us to spot an error in the simulations and Théo Dessertaine, Nils Ottink, and Nathan Coste for their careful revision of the paper.
We appreciate Elina Kostanian's assistance in enhancing the visual rendering of the figures. Live long and prosper.
\inlinegraphics{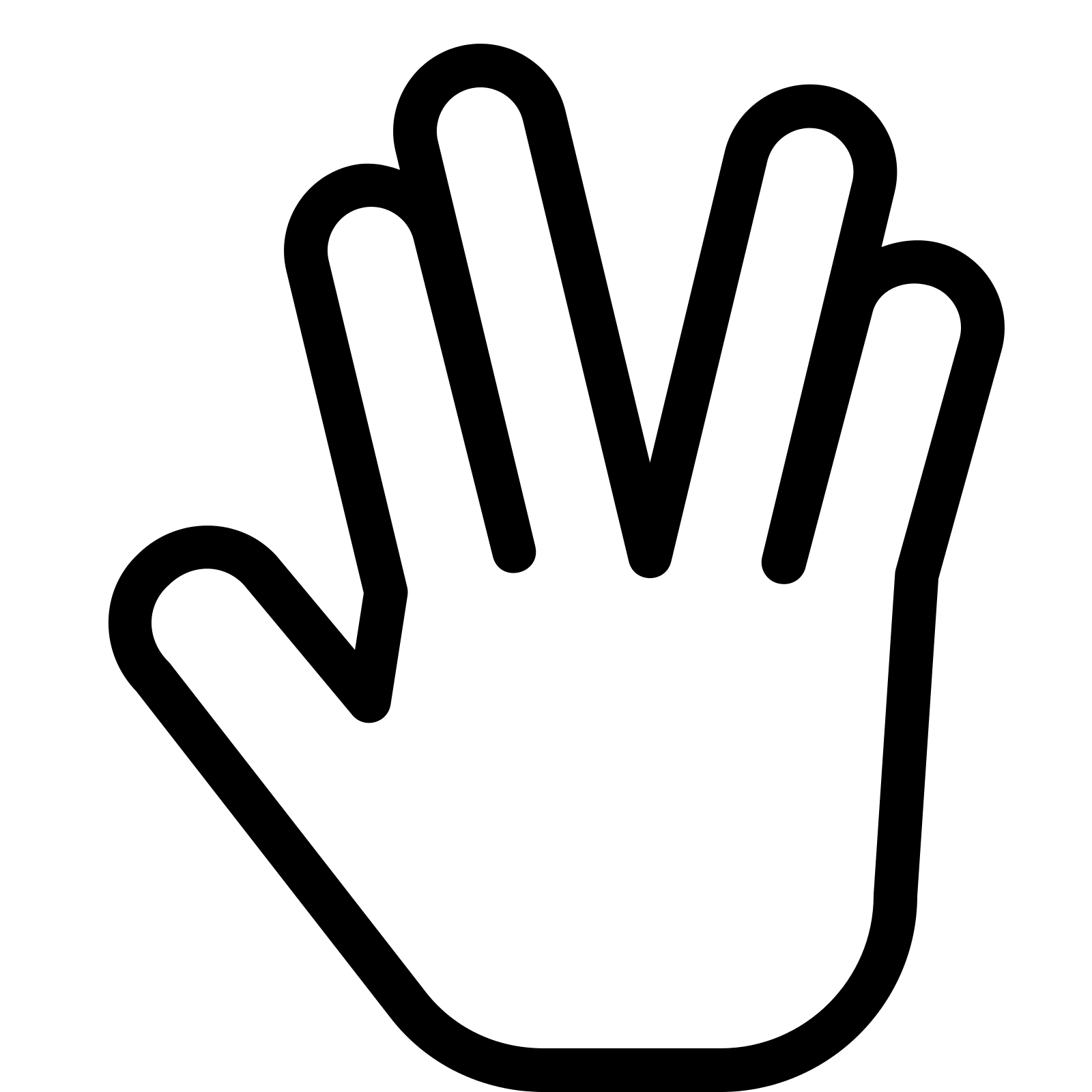}   
This work has been co-funded by the European Commission as part of the EIC accelerator program under the grant agreement 190188855 for SEPOQC project, by the Horizon-CL4 program under the grant agreement 101135288 for EPIQUE project, and by the CIFRE grant n°2022/0532.
\end{acknowledgments}

\section*{Author contributions}
G.G.\ and P.H.\ contributed equally to this work.
P.H.\ proposed the main architecture ideas, wrote the paper and contributed to the visualization and to the simulations. 
G.G.\  contributed to the main architecture ideas, to visualization, to simulation, to research on the decoder and to the paper writing.
P-E.E.\ wrote the paper, contributed to visualization and to the simulations and participated to conceptualization. 
S.C.W.\ contributed to the physical noise models and edited the paper. 
A.S.\ contributed to research on the decoder and discussions on the code and draft.
S.M.\ supervised.

\appendix

\section{Discussion on the scalability of SPOQC with non-local LDPC codes}
\label{subsec_nonlocal_ldpc}

Strictly speaking, implementing a non-local LDPC code using this architecture cannot be scalable since it requires quantum channels that increase with the computer's size, and thus increase time latencies and fiber losses. Long-range code could significantly reduce the resource overhead~\cite{gottesman2013fault, fawzi2020constant, cohen2022low}, so we should still explore them. As long as the time latencies and the fiber loss do not become the principal source of errors, such FTQC architecture could have a practical interest. 

Relaxing this requirement is sound as long as the overhead it causes is small compared to other sources of errors. 
For fiber loss, given that a qubit module has a spatial footprint of $V$, and that we can use a $D$-dimensional qubit arrangement ($D\leq 3$ for a spatial layout compatible with our space-time), the losses scales in principle with $\propto e^{- N^{1/D} (V^{1/3} / L_{\rm att})}$ with $N$ the number of qubits in the FT quantum computer and $L_{\rm att}$ the channel attenuation length. This typical loss should be smaller than the loss thresholds, which is all the easier to achieve for platforms with spatial footprint $V \ll {L_{\rm att}}^3$. For telecom photons, the typical attenuation distance of a fiber is of the order of $22$ km, and we can also envision free-space optical links if need be~\cite{litinski2022active}.

Time latencies can also be challenging to deal with. However, every hardware is limited by communication delays, and photonics offers an advantage in executing non-local gates. In the end it boils down to a similar analysis as the latency scales with $\propto k N^{(1/D)} V^{(1/3)}$, and again the typical latency time should be
much smaller than the  percentage threshold of the coherence time $T_2$ of the spin qubits.

\section{Detail on RUS gate channels}
\label{RUS_channels}
We detail the behavior of the success and repeat channels of the RUS gate displayed in Table~\ref{tab:RUS_sumup} on stabilizers. There are two spin qubits and two photonic qubits involved that we will label $\rm qe 1\rm /p h1$ and $\rm qe 2/\rm ph 2$.
We investigate the effect of the channels by evaluating their transformation on input stabilizer operators.
For example, here is the transformation of a CZ gate, namely the one we expect from the success of the RUS gate.
$$ \begin{array}{ccc}
 \text{Input} && CZ \\
\begin{array}{|ccc|}
         & \rm qe 1 & \rm qe 2\\  \hline\hline
        +&X&I\\
        +&I&X\\
        +&Z&I\\
        +&I&Z \\ \hline
    \end{array} & \rightarrow &
    \begin{array}{|ccc|}
         & \rm qe 1 & \rm qe 2\\  \hline\hline
        +&X&Z\\
        +&Z&X\\
        +&Z&I\\
        +&I&Z \\ \hline
    \end{array}
    \end{array}$$ 
\paragraph{Stabilizers after photon emissions.}
The emission operator of the quantum emitter is assumed to be 
$$
E_{\rm qe, ph} = \ket{0}_{\rm qe} \ket{0}_{\rm ph} \bra{0}_{\rm qe} + \ket{1}_{\rm qe} \ket{1}_{\rm ph} \bra{1}_{\rm qe}.
$$
This is equivalent to preparing the photon qubit in $\ket{0}_{\rm ph}$ and then applying a CNOT between the spin and the photonic qubits. Hence, independently of the spin state, the spin and photonic qubits are stabilized by $Z_{\rm qe}Z_{\rm ph}$.
If the spin qubit is stabilized by $Z_{\rm qe}$ before emission, its stabilizer is preserved upon emission. However, if the spin is stabilized by $X_{\rm qe}$, then the stabilizer becomes $X_{\rm qe}X_{\rm ph}$. Here are the transformation of the  four spin stabilizers resulting from two emissions followed by the two photon stabilizers that appear independently of the spins' states.
$$
\begin{array}{|ccccc|}
         & \rm qe 1 & \rm qe 2 & \rm ph 1 &\rm ph 2\\ \hline\hline
        +&X&I&X&I\\
        +&I&X&I&X\\
        +&Z&I&I&I\\
        +&I&Z&I&I\\ \hline 
        +&Z&I&Z&I\\
        +&I&Z&I&Z\\ \hline
    \end{array}
$$

\paragraph{Repeat measurement channel.}
The ``repeat'' outcome corresponds to measuring $X_{\rm ph 1} $ and  $X_{\rm ph 2}$. In that case, their outcomes $m_1$ and $m_2$ always verify $m_1 = m_2$. The measurement yields
$$\begin{array}{|ccccc|}
         & \rm qe 1 & \rm qe 2 & \rm ph 1 &\rm ph 2\\ \hline \hline
       (-1)^{ m_1}&X&I&I&I\\
        (-1)^{ m_1}&I&X&I&I\\
          +&Z&I&I&I\\
          +&I&Z&I&I\\  \hline
          (-1)^{ m1}&I&I&X&I\\
          (-1)^{ m1}&I&I&I&X\\ \hline
    \end{array}.
$$

One can note that the $Z_{\rm qe 1}Z_{\rm qe 2}$ conditioned on $m_1= 1 $  would flip the sign of both $X_{\rm qe 1}$ and $X_{\rm qe 2}$, making the gate return the identity gate systematically. This ensures that no indirect measurement is performed on the spin qubits.
\paragraph{Successful measurement channel.} The success happens when there is one photon per pair of modes at the detection. This corresponds to  $X_{\rm ph 1}X_{\rm ph 2}$ and $Z_{\rm ph 1}Y_{\rm ph 2}$ measurements. The Pauli $Y$ appears with the phase shifter on mode $3$. The measurements induce 
$$\begin{array}{|ccccc|}
         & \rm qe 1 & \rm qe 2 & \rm ph 1 &\rm ph 2\\ \hline\hline
        +&Y&Z&Y&Z\\
        +&Z&Y&Z&Y\\
        +&Z&I&I&I\\
        +&I&Z&I&I\\ \hline 
        +&I&I&X&X\\
      (-1)^{ m_2}&I&I&Z&Y\\ \hline
    \end{array}.
$$

Upon rewriting of the tableau, we have 
$$\begin{array}{|ccccc|}
         & \rm qe 1 & \rm qe 2 & \rm ph 1 &\rm ph 2\\ \hline\hline
     (-1)^{ 1+m_2}&Y&Z&I&I\\
      (-1)^{ m_2}&Z&Y&I&I\\
        +&Z&I&I&I\\
        +&I&Z&I&I\\ \hline 
     (-1)^{ 1+m_2}&I&I&Y&Z\\
      (-1)^{ m_2}&I&I&Z&Y\\ \hline
    \end{array}.
$$

Since $S Y S^\dag = - X$, we have that the corrections $S_{qe1} {S_{qe2}}^\dag$ if $m_2 = 0$ and ${S_{qe1}}^\dag {S_{qe2}}$ if $m_2=1$ allow us to recover exactly a CZ operation on the spin qubits.

\section{Physical noise models}
\label{sec_physical_noise}

\subsection{Realistic modeling of the relaxation and dephasing of a quantum emitter}
A quantum emitter can be seen as a quantum memory (its spin), that can be interfaced with photons. However, due to its interaction with its environment, this spin qubit cannot store quantum information for an infinitely long time. In practice, due to its interaction with the environment, it is limited by its relaxation time and its coherence time, respectively denoted $T_1$ and $T_2$.

Let $\hbar\omega_{\uparrow\downarrow}$ be the energy splitting between the lower-energy $\ket{\downarrow}$ spin state and the higher-energy $\ket{\uparrow}$ spin state. If $\hbar\omega_{\uparrow\downarrow}$ is much larger than the temperature of the environment $k_B T$, then the qubit will relax to the state $\ket{\downarrow}$. However, in the opposite limit $\hbar\omega_{\uparrow\downarrow}\ll k_B T$, the qubit will relax towards a completely mixed spin state. A simple model of qubit relaxation due to a thermal bath is described by a Lindblad master equation $d\rho(t)/dt = \mathcal{L}\rho(t)$ \cite{breuer2002theory} where the generator is
\begin{equation}
\begin{aligned}
    \mathcal{L}\rho = &-\frac{i}{\hbar}\left[H,\rho\right] + \gamma_0(n_\text{th}(T)+1)\mathcal{D}({\sigma}_-)\rho\\ &+ \gamma_0n_\text{th}(T)\mathcal{D}({\sigma}_+)\rho(t) + \frac{\gamma^\star(T)}{2}\mathcal{D}({\sigma}_z)\rho
\end{aligned} \vspace{0.1cm}
\end{equation}
and where $H=\hbar\omega_{\uparrow\downarrow}{\sigma}_z/2$ is the spin qubit Hamiltonian, $\sigma_z=\ket{\uparrow}\bra{\uparrow}-\ket{\downarrow}\bra{\downarrow}$ is the Pauli-Z operator, $\sigma_-=\ket{\downarrow}\bra{\uparrow}$ is the spin lowering operator, $\sigma_+=\ket{\uparrow}\bra{\downarrow}$ is the spin raising operator, $\gamma_0$ is the zero-temperature relaxation rate defined by the spin-bath coupling strength, $n_\text{th}(T)= 1/(e^{\hbar\omega_{\uparrow\downarrow}/k_BT} - 1)$ is the thermal population of the bath, and $\gamma^\star(T)$ is the spin pure dephasing rate at temperature $T$. The non-Hermitian evolution is governed by the Lindblad dissipator defined by
\begin{equation}
    \mathcal{D}(L)\rho = L\rho L^\dagger - \frac{1}{2}L^\dagger L \rho - \frac{1}{2}\rho L^\dagger L.
\end{equation}

The noise channel $C_t$ acting on the qubit for a duration of time $t$ is the general solution to the Lindblad master equation: $dC_t/dt =\mathcal{L}C_t$ in the frame rotating with the spin qubit precession. Since the generator $\mathcal{L}$ is time-independent, the solution is given by $C_t = e^{t\mathcal{L}}$, which can be analytically solved by diagonalizing $\mathcal{L}$ to obtain
\begin{widetext}
    
\begin{equation}
\begin{aligned}
    C_t(\rho) &= \frac{1}{(\gamma_{\downarrow}+\gamma_{\uparrow})}\left[\left(\gamma_{\downarrow}+\gamma_{\uparrow}e^{-t(\gamma_{\downarrow}+\gamma_{\uparrow})}\right)\rho_{\uparrow\uparrow}
    \rho\rho_{\uparrow\uparrow}+\left(\gamma_{\uparrow}+\gamma_{\downarrow}e^{-t(\gamma_{\downarrow}+\gamma_{\uparrow})}\right)\rho_{\downarrow\downarrow}
    \rho\rho_{\downarrow\downarrow}\right.\\
    &\hspace{10mm}\left.+\gamma_{\uparrow}\left(1-e^{-t(\gamma_{\downarrow}+\gamma_{\uparrow})}\right)\rho_{\downarrow\downarrow}
    \rho\rho_{\uparrow\uparrow}+\gamma_{\downarrow}\left(1-e^{-t(\gamma_{\downarrow}+\gamma_{\uparrow})}\right)\rho_{\uparrow\uparrow}
    \rho\rho_{\downarrow\downarrow}\right.\\
    &\hspace{10mm}\left.(\gamma_{\downarrow}+\gamma_{\uparrow}) e^{-t((\gamma_{\downarrow}+\gamma_{\uparrow})/2-\gamma^\star)}\left(\rho_{\downarrow\uparrow}
    \rho\rho_{\uparrow\downarrow}+\rho_{\uparrow\downarrow}
    \rho\rho_{\downarrow\uparrow}\right)\right]
\end{aligned}
\end{equation}
\end{widetext}

where $\rho_{ij} = \ket{i}\bra{j}$, $\gamma_{\uparrow} = \gamma_0 n_\text{th}(T)$, and $\gamma_{\downarrow} = \gamma_0(n_\text{th}(T) + 1)$. This channel is not a Pauli channel in general. However, in the limit that $n_\text{th}\gg 1$ (or where $\hbar\omega\ll k_B T$), we have $\gamma_{\uparrow}\simeq \gamma_{\downarrow}=\gamma$. In this case, we can write
\begin{equation}\label{eq_spin_decoh}
    \begin{aligned}
    C_t(\rho) & = (1-p_X-p_Y-p_Z)\rho \\ & + p_XX\rho X + p_Y Y\rho Y + p_Z Z\rho Z    
    \end{aligned}
\end{equation}
where
\begin{equation}
\begin{aligned}
    p_X &= p_Y = \frac{1-e^{-2t\gamma}}{4}\\
    p_Z &= \frac{1-e^{-t(\gamma+\gamma^\star)}}{2} - \frac{1-e^{-2t\gamma}}{4}.
\end{aligned}
\end{equation}
This is a Pauli error channel where the effective relaxation time is $T_1 = 1/2\gamma$, the coherence time is $T_2 = 1/(\gamma+\gamma^\star)$, and we have that $T_2\leq 2 T_1$.

\subsection{Partial photon distinguishability}
\label{subsec_distinguishability}

\begin{figure}[!ht]
    \centering
    \includegraphics[width=0.9\columnwidth]{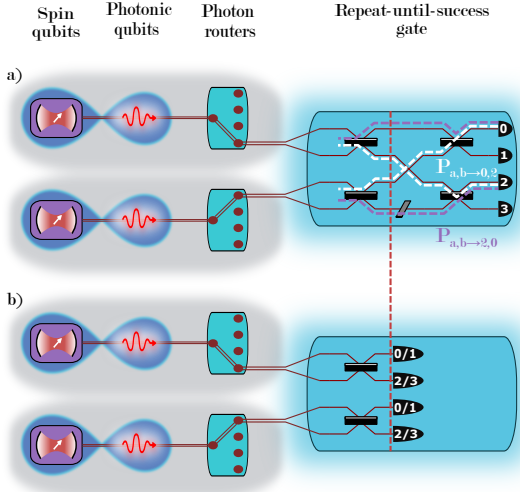}
    \caption{(a) Linear-optical interferometer with input spin quantum emitters. The paths $P_{a,b \to 0,2}$ and $P_{a,b \to 2,0}$ are represented in white and violet dashed lines respectively. The vertical red dashed line corresponds to the zone before the mode swap mentioned in the main text. (b) ``Equivalent'' linear-optical interferometer for distinguishable photons.}
    \label{fig_distinguishability}
\end{figure}

Photons emitted by two different quantum emitters should interfere together on a linear-optical interferometer. These photons should thus be perfectly indistinguishable to lead to a unity fidelity RUS gate. The photon indistinguishability is denoted by $M$ and is equal to 1 for perfectly indistinguishable photons, and 0 for completely distinguishable photons (which cannot interfere together). This quantity corresponds to the overlap of the wavefunctions of two single photons and can be calculated experimentally with a Hong-Ou-Mandel experiment~\cite{hong1987measurement}. We are interested in the partial distinguishability $D = 1 - M$, corresponding to an error afflicting our system. For simplicity, we consider that this distinguishability afflicts identically all the photons emitted by all the quantum emitters.

Having photon indistinguishability is detrimental to the RUS gate, so we want to evaluate its impact. Physically, it forbids the erasure of the photon ``which-path''  information at the core of linear-optical gates.
A path corresponds to the trajectories of photons into the interferometers to the detectors. For example ``photons $a$ and $b$, respectively emitted by quantum emitters $a$ and $b$, are respectively detected in mode $0$ and $2$'' is a path, that we call path $P_{a,b \to 0,2}$, see Fig.~\ref{fig_distinguishability}(a).

We can have a second path $P_{a,b \to 2, 0}$ ``photons $a$ and $b$, respectively emitted by $a$ and $b$, are respectively detected in mode $2$ and $0$''.
These two paths are not identical, $P_{a,b \to 0, 2} \neq P_{a,b \to 2,0}$, but have the same detection pattern $(0, 2)$.
For indistinguishable photons, given the detection pattern $(0, 2)$, it is physically impossible to know which path the photons have followed, hence the name of which-path information erasure, which can lead to interference between these paths and linear-optical entangling gates. 
However, If we have distinguishable photons, it is theoretically possible to identify which detector has detected photon $a$. In this example, we could know the photons have followed path $P_{a,b \to 0,2}$ or path $P_{a,b \to 2,0}$ with that knowledge. While we don't have access to this information, this leads to the absence of an entangling gate and a statistical mixture of these two paths.

For distinguishable photons, if we were able to discriminate them at the detector, we could easily retrieve their path up to before the mode swapping (see the dashed red line in Fig.~\ref{fig_distinguishability}(a)). Therefore, the distinguishable case corresponds to the linear-optical interferometer in Fig.~\ref{fig_distinguishability}(b), which performs a photon $X$ measurement in the photonic qubit basis.

\paragraph{Repeat detection pattern}
For partial failure detection patterns, the effect on the spin is similar whether or not we have distinguishable or indistinguishable photons. Indeed, we can identify which path each photon has followed in a partial failure, so there is no which-path information erasure. Therefore, we should only focus on the successful detection patterns. We can also obtain the detection patterns $(0,1)$ and $(2,3)$ with distinguishable photons. The Hong-Ou-Mandel interference (on the last beamsplitters of the linear-optical interferometer) forbids these detection patterns with indistinguishable photons. For the same reasons as before, these patterns are well handled and do not lead to additional errors induced by distinguishable photons.

\paragraph{Successful detection pattern}
For indistinguishable photons, successful detection patterns perform a CZ gate on the spin qubits up to some correction unitary $C$ that depends on the detection pattern. We focus on the detection pattern $(0,2)$, for which $C = S_a {S_b}^\dag$, but we can apply the same reasoning to any successful detection pattern.
For distinguishable photons, following path $P_{a,b \to 0, 2}$ (respectively $P_{a,b \to 2, 0}$) corresponds to a measurement of the photons in the $\ket{+_a, -_b}$ state (respectively $\ket{-_a, +_b}$). This corresponds to applying a unitary $Z_b$ (respectively $Z_a$) on the spin states. However, in practice, we don't know that the photons were distinguishable, and we don't know which path has been followed. Therefore, we perform the correction $C$ on a statistical mixture of the two paths. Therefore, for partially distinguishable photons, we perform the channel:
\begin{align}
C_D(\rho)& = (1 - D) \text{CZ}_{a,b} \rho \text{CZ}_{a,b} \nonumber\\ & + \frac{D}{2} S_a {S_b}^\dag (Z_a \rho Z_a + Z_b \rho Z_b) {S_a}^\dag S_b.  \nonumber  
\end{align}

We can rewrite this as the desired gate channel, followed by the error channel:
$$
C_D(\rho) = C_{D, err}( \text{CZ}_{a,b} \rho \text{CZ}_{a,b})
$$
with
\begin{align}
C_{D, err}(\rho)&  = (1-D) \rho \nonumber + D \text{CZ}_{a,b} \tilde \rho \text{CZ}_{a,b}. \nonumber  
\intertext{with}
\tilde \rho & = S_a S_b^\dagger ( \frac{Z_a\rho Z_a}{2}  + \frac{Z_b\rho Z_b}{2} )S_a^\dagger S_b \nonumber
\end{align}

Because, this channel is not a Pauli error channel, we simplify it by replacing the noise with a completely dephasing channel:

$$
C_{D, err}(\rho) \to C_{\rm RUS, s}(\rho) = (1-D) \rho + D C_{Z_a}  C_{Z_b} (\rho)
$$ 
with
$$
C_{Z_i}(\rho) = \frac{1}{2} (\rho + Z_i \rho Z_i).
$$
Indeed the channel $\rho \mapsto\frac{Z_a\rho Z_a}{2}  + \frac{Z_b\rho Z_b}{2}$
acts as $$\rho \mapsto 
\begin{pmatrix}
    \rho_{11}&0&0&\rho_{14}\\
    0&\rho_{22}&\rho_{23}&0\\
    0&\rho_{32}&\rho_{33}&0\\
    \rho_{41}&0&0&\rho_{44}
\end{pmatrix},
$$
whereas the channel $\rho \mapsto C_{Z_a}  C_{Z_b} (\rho)$ acts as
$$\rho \mapsto 
\begin{pmatrix}
    \rho_{11}&0&0&0\\
    0&\rho_{22}&0&0\\
    0&0&\rho_{33}&0\\
    0&0&0&\rho_{44}
\end{pmatrix}, $$
making the density matrix diagonal, hence insensitive to CZ and $S$ gate channels.
Overall, we assume more erasure than in reality and we overestimate the effect of distinguishability.

\subsection{Errors involved in the SPOQC architecture}
For this work, we have decided to focus on photon loss, photon distinguishability and spin decoherence time, which we consider to be the dominant errors for an implementation of our architecture using quantum dot spins. Obviously, other noise mechanisms will likely be present in this architecture and should also be taken into account for a practical implementation. In the following, we motivate this choice, consider different types of common errors and the current performances reached experimentally with quantum dot platforms. 

\paragraph{Spin errors}
Regarding spin errors, we center our analysis on the spin decoherence time rather than evaluating single-qubit operation error. This is crucial since a RUS gate takes a finite time and should be compared to a relevant timescale during which we can keep quantum information in a physical qubit. The coherence time is also critical, since it used to be a limiting factor for spins in quantum dots. Yet, a recent work~\cite{zaporski2023ideal} managed to increase it to 100~\textmu s timescale. Excluding other errors, a complete RUS gate should thus last at most $t_{\rm rus} \leq 2$\textmu s. Other types of spin manipulation errors such as the spin qubit initialization, gate, and measurement noises should be also accounted for in a more complete analysis.

\paragraph{Single-photon impurity}
A typically photonic error is single-photon impurity, i.e. the noise corresponding to a quantum emitter generating two photons (or more) instead of one. In a preliminary analysis, we could expect this error to have a similar effect as photon loss. Indeed, the ``success'' and ``repeat'' outcomes of a RUS gate trial correspond to having exactly two detected photons. Thus, having one photon less (i.e. photon loss) or more (i.e. single-photon impurity) leads to a similar failure outcome. Therefore, we should expect a photon impurity threshold at least as high as the photon loss threshold (when not considering other sources of error)~\footnote{Actually, the threshold could be even higher since some 3-detected-photon pattern in the RUS gate could unambiguously lead to a repeat or success outcome for the RUS gate. This is not the case for loss.}.

Nevertheless, a more complete analysis is needed if we consider single-photon impurity together with loss since the joint effect of these errors becomes unclear: for example, when two photons are emitted by the same quantum emitter and one of them is lost, the detected signal is ambiguous for the RUS gate. Since the quantum dot sources we generate already have high purity, we focus our analysis onto photon loss, which is currently much more critical for these sources. For example, sources with single-photon impurity of less than $p_2 \leq 0.4\%$ (with $p_2$ the probability of emitting 2 photons) have been demonstrated~\cite{claudon2010highly} while the current best single-photon source still has a photon loss $1 - \eta$ at collection of $29\%$\cite{ding2023high}.

\paragraph{Photon distinguishability}
Similarly, we emphasize that we are considering the distinguishability between photons originated from different sources. This is significantly different from having excellent indistinguishability between photons coming from the same source, which have already been demonstrated~\cite{somaschi2016near}. For different sources, the smallest photon distinguishability demonstrated with quantum dots so far is D=7\% ~\cite{zhai2022quantum}. We also point out that our scheme exploits two transitions of a quantum dot, which should thus be identical on all the degrees of freedom except the one onto which we encode the photonic qubit.

\paragraph{Full set of noise mechanisms}
Taking all the above into account, a complete analysis of the architecture under a large set of noise mechanisms is necessary but beyond the scope of this work which focuses on the architecture presentation and a performance assessment under the main source of noise. A more complete error analysis is left to future work.

\section{Notes on the numerical simulations}
\label{sec_notes_simu}

\subsection{Simulation details}

\begin{figure}[ht]
    \centering
    \includegraphics[width=.9\columnwidth]{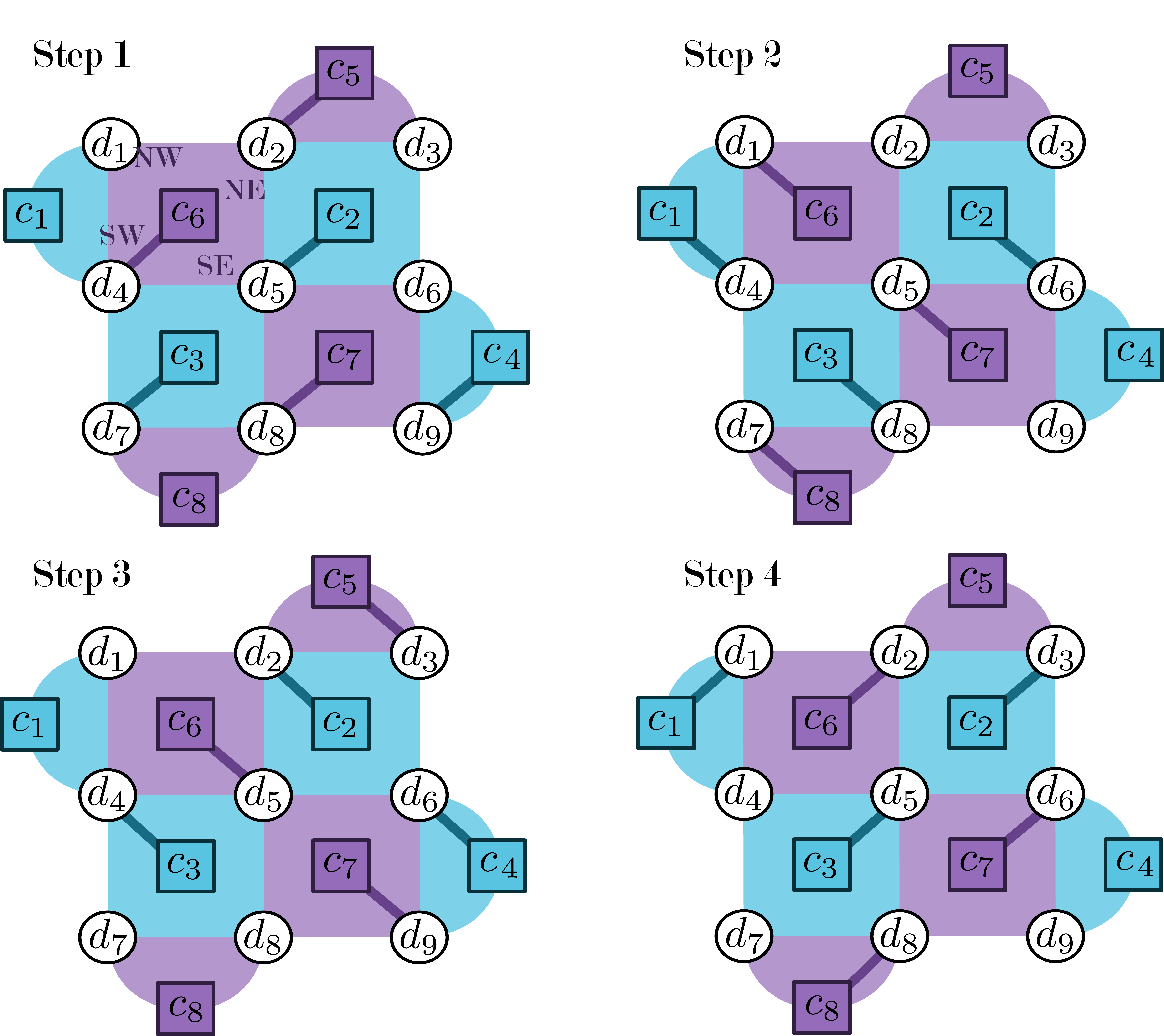}
    \caption{Illustration of the order in which the RUS CZ are performed. Thick edges corresponds to a CZ gate being performed at a given step.}
    \label{fig_order_cz}
\end{figure}

Here, we give some details about how we perform the numerical simulations.
We used Stim~\cite{gidney2021stim} to perform efficient stabilizer simulations and Pymatching~\cite{higgott2022pymatching, higgott2023sparse} for the minimum-weight perfect matching decoder. 

To measure the stabilizers, we use a noisy indirect stabilizer measurement circuit as was presented in Fig.~\ref{fig:indirect_meas_rus}. However, instead of performing each stabilizer measurement one after the other, we do them ``all at once''. It avoids the unnecessary propagation of errors. We use the strategy proposed in~\cite{mcewen2023relaxing} and illustrated in  Fig.~\ref{fig_order_cz}.
The edges connecting a check qubit are oriented along the North West (NW, e.g. edge $(c_6, d_1)$), North East (NE, e.g. $(c_6, d_2)$), South East (SE, e.g. $(c_6, d_5)$) and South West (SW, e.g. $(c_6, d_4)$) directions. We perform the CZ gates in four steps illustrated in Fig.~\ref{fig_order_cz}: for $X$-type (respectively $Z$-type) stabilizers, we start by SW, then NW, SE, and finally NE (respectively, SW, SE, NW, and then NE) CZ gates.
This way a complete cycle of syndrome measurements only takes the time of four entangling steps and does not depend on the distance $d$ of the rotated surface code.

Instead of a theoretical CZ gate, we perform a RUS CZ gate, which is a heralded gate and can either succeed, fail, or abort. We consider the abort and failure cases with the same error channel $C_{\rm RUS, f}(\rho)$ in the simulations. This channel corresponds to the failure case, while there is a priori no phase erasure in the case of the abort case. Therefore, this assumption leads to an error overestimation in the aborted case. Yet, it also simplifies the analysis as discussed in the main manuscript, see \ref{sec_results}.

\paragraph*{Heralded errors.}
Stim did not account for heralded errors natively when we perform the simulations.
To account for heralded gate in Stim, we consider using an extra ancilla qubit $c$. For a CZ gate between qubit $a$ and $b$, we use the ancilla qubit $c$, initialized in the $\ket{0}$ state, as:
  
\begin{widetext}
  
\begin{equation}
C_{\rm RUS}(\rho_{ab} \otimes \op{0}_c) = (1 - p_F) C_{\rm RUS, s}(\rho_{ab}) \otimes \op{0}_c + p_F C_{RUS, f}(\rho_{ab}) \otimes X_c\op{0}_c X_c.
\label{eq_results_descriptions}
\end{equation}

We then measure qubit $c$.
If it was not flipped, the successful gate was applied. Otherwise, $X_c \op{0}_c X_c = \op{1}_c$ , which means we applied the unsuccessful gate.

Moreover, since $C_{RUS, f}(\rho)$ is the two-qubit completely dephasing channels,
$C_{RUS, f}  C_{\rm RUS, s}(\rho) = C_{\rm RUS, f}(\rho)$, therefore we can rewrite Eq.~\eqref{eq_results_descriptions} as

\begin{align}
C_{\rm RUS}(\rho_{ab} \otimes \op{0}_c) =  C_{\rm RUS, f, err}  (C_{\rm RUS, s}(\rho_{ab}) \otimes \op{0}_c) \nonumber    
\end{align}
with
\begin{align}
C_{\rm RUS, f, p_F} (\rho_{ab} \op{0}_c) = (1 - p_F) \rho_{ab} \otimes \op{0}_c \nonumber + p_F C_{\rm RUS, f}(\rho_{ab}) \otimes \op{1}_c.  
\end{align}
\end{widetext}
This ``hack'' enables the construction of circuits involving heralded errors in Stim.~\footnote{This solution was provided by Craig Gidney in \url{https://quantumcomputing.stackexchange.com/questions/26582/how-do-i-perform-an-erasure-error-in-stim}}. The use of the ``CORRELATED\_ERROR'' and ``ELSE\_CORRELATED\_ERROR'' commands in stim does not mean errors to be decoded are correlated. This is the mechanism used to heralds errors. Here measuring qubit $c$ in $\ket{1}$ heralds the dephasing of both qubit $a$ and $b$. 

\paragraph*{Decoding heralded errors.}
We used Pymatching's minimum weight perfect matching (MWPM) decoder \cite{higgott2022pymatching}.
The latter works together with Stim \cite{gidney2021stim} that determines a detector error model, namely the error model on the check operators measured during the simulations. Pymatching builds a matching graph out of it and applies the MWPM algorithm. This Stim/Pymatching interfacing allows tracking errors efficiently and automatically, even with complex error models.
However, this interfacing does not account for heralded errors natively.
To address this issue, we update manually the matching graph to be decoded with the heralded errors of each run.
While this method of handling errors may not be the most efficient, it is sufficient for our current needs and highly versatile. It can theoretically be applied to any type of heralded uncorrelated errors. Moreover, it accommodates partial erasure, potentially finding practical applications in other platforms where similar strategies can be employed \cite{Suchara2015Leakage}.

This method requires the initial error model of the circuit without any heralded error and a register of the heralded error models of each herald. When running the circuit, the herald signal determines which heralded error models are to be incorporated into the initial error model. In term of matching graph, this corresponds to updating the graph accordingly as was done in Ref.~\cite{stace2010error}. Predicting RUS-heralded error models is more challenging than predicting simple qubit erasures, therefore mock circuits are simulated in order to extract each heralded error models. One can observe that due to the construction of the RUS-heralded errors, the resulting error models will be made up of uncorrelated errors. This allows for a proper use of the MWPM algorithm.

\paragraph*{Spin dephasing.}
The last type of errors accounted for in our simulations are spin decoherence errors that lead to a continuous time error $C_t(\rho)$ from Eq.~\eqref{eq_spin_decoh}. Assuming $T_2 \ll T_1$, and taking $T_1 \to \infty$, this channel has the form of a dephasing channel:
$$
C_t(\rho) = (1 - p_Z) \rho + p_Z Z\rho Z.$$
with $p_Z$ depending on $t$.
We consider all the decoherence to occur for a time $t= t_{\rm RUS}$ before each of the four time steps discussed previously. We can do this since the spin decoherence channel commutes with all the RUS gate channels. 
A complete error correction cycle thus lasts $4t_{\rm RUS}$ if we assume that other operations take negligible time compared to a RUS gate.

We also consider that we can perform any single-qubit operation and measurement with unit fidelity. We leave the study of more complete error models to future work. 

\subsection{Numerical results for independent errors}

In all the numerical simulations, we evaluate the error threshold by performing a fault-tolerant initialization of the logically-encoded qubit. For a $\llbracket d^2, 1, d \rrbracket$ rotated surface code, we start with $n$ disentangled physical qubits in the state $\ket{0}^{\otimes n}$ (respectively $\ket{+}^{\otimes n} = (\ket{0} + \ket{1})^{\otimes n}$) and we want to encode a logical $\ket{0}_L$ (respectively $\ket{+}_L$) qubit using noisy gates and qubits. We do so by performing $d$ cycles of error correction. We then measure all the qubits in the $Z$ (respectively $X$) basis to evaluate whether or not the fault-tolerant initialization was successful: if the measurement gives $\ket{1}_L$ (respectively $\ket{-}_L = \ket{0}_L - \ket{1}_L$), we have a logical error. 

\subsection{Determination of the fault-tolerant region of errors}

We wish to explore a tri-dimensional parameter space corresponding to the three main physically relevant errors present in the SPOQC architecture: photon loss, partial distinguishability, and spin decoherence time. We want to extract the border between the FT and the non-FT regimes. This border corresponds to a surface $S$ in this tri-dimensional space.

We already know three points from $S$ consecutive to the study of individual errors: $\overrightarrow{OA} = (p_{\rm F, th, 0}, 0, 0)$, $\overrightarrow{OB} = (0, t_{\rm th, 0}, 0)$, and $\overrightarrow{OC} = (0, 0, D_{\rm th, 0})$.

To reduce the parameter space exploration, we make the assumption that the border $S$ is ``not too far'' from the plane $\hat S$ passing by these three points and defined by the equation $\frac{x}{p_{\rm f, th, 0}} + \frac{y}{t_{\rm th, 0}} + \frac{z}{D_{\rm th, 0}} = 1$. We verify that this assumption is sound later on through the simulations. 

To find the correct surface $S$, we first start by taking a tessellation of points on the plane $\hat S$ (more precisely on the intersection of this plane with ${\left(\mathbb{R}^+ \right)}^{3}$). For each point $M$, we consider the line $\ell_M$ passing through the origin $O$ defined by the vector $\overrightarrow{OM}$ and want to identify the intersection point $N$ of $\ell_M$ with the real FT-threshold surface $S$. 
The line $\ell_M$ is parameterized by $\overrightarrow{OM'}(w) = w \overrightarrow{OM}$, and we look for the value $w_{\rm th}$ corresponding to the intersection of $\ell_M$ with the surface $S$. We do so by performing simulations for combined errors corresponding to $w \overrightarrow{OM}$ on the error space. The error intensities all increase with $w$, and we identify the FT threshold corresponding to the crossing points between two curves of distance $d$ and $d+2$ as a function of $w$. We use $d= 11$ in our simulations. This crossing point is $w_{\rm th}$. It is the ``real'' crossing point between $\ell_M$ and $S$. The assumption that $S$ is close to the planar surface $\hat S$ helps us in using a small interval $[w_{\rm min}, w_{\rm max}]$ of values close to $1$ to avoid exploring the full error space. After a first refinement, we use an interval $[w_{\rm min}, w_{\rm max}] = [0.85, 1]$ and ensure that the assumption is valid by checking that $w_{\rm th}$ is always found within $[w_{\rm min}, w_{\rm max}]$. To obtain the border $S$ in Fig.~\ref{fig_RUS_threshold}, we plot the threshold point that we have found along each line $\ell_M$ of the tessellation of $n_p = 120$ points.

\section{Variant with hybrid RUS gates}
\label{app_hybrid_RUS}

The RUS gate offers the best robustness against photon loss, but as we can see in Fig.~\ref{fig_physical_threshold}, the need of using many trials causes stringent requirements on the spin coherence time.
A RUS gate is indeed relatively long as it requires to store quantum information on the spins for $k$ trials corresponding to the time of photon emission, transfer, and detection and also from the classical processing and feedforward delays.  

\begin{figure}[ht]
    \centering
    \includegraphics[width=1\linewidth]{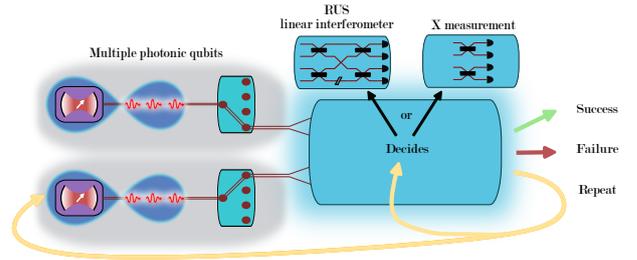}
    
    \caption{Hybrid RUS gate including spin qubits capable of emitting multiple entangled photons, an adaptive interferometer and detectors. The linear-optical interferometer is adaptive and depends on previous measurement outcomes. Horizontal black rectangles correspond to $50:50$ beamsplitter, the inclined parallelogram corresponds to a $-\pi/2$ phase shifter, numbered semiovals are photon-number-resolving detectors.}
    \label{fig_rus_boosted_gates}
\end{figure}

An alternative strategy relies on boosted fusion gates~\cite{hilaire2023linear} and is potentially much faster as it requires only one trial step. This gate corresponds to having each spin sending sequentially $n$ photons to an adaptive linear-optical interferometer. This scenario differs from RUS gates, where we only emit one photon per trial. The linear-optical interferometer depends on the previous measurement outcomes. If no previous measurements has yielded a successful detection pattern, it corresponds to the RUS gate. Otherwise, it corresponds to two independent $X$ measurements on the photonic qubits. Fig.~\ref{fig_rus_boosted_gates} illustrates a physical implementation of such a gate.  
We can picture this as making the $k=n$ trials of an equivalent RUS gate ``all at once'', that is, $n$ trials of the RUS gate is implementing by a hybrid RUS gates with $n$ photons per trial, provided that no photon is lost.

Interestingly, all the results derived in the paper are valid if we replace RUS gates by hybrid RUS gates. It means replacing the building block based on single-photon emission with a hybrid gate based on multiple sequential photon emissions. 
If we consider this hybrid scheme, hereafter denoted as HRUS($k,n$) for at most $k$ trials with $n$ photons per trial, we can increase the fault-tolerant region of our architecture.

HRUS($k,n$) is successful if we obtained at least one correct detection pattern \emph{and} if all the emitted photons have been successfully detected:
$$p_s = (1 - 2^{-n}) (\eta_a \eta_b)^{n}.$$
The hybrid gate has failed if at least one photon is lost:
$$p_f = 1 - (\eta_a \eta_b)^{n},$$
and it has been aborted otherwise:
$$p_r = 2^{-n} (\eta_a \eta_b)^{n}.$$
In that case, the total success, failure and abort probabilities are given by
\begin{equation*}
    \begin{aligned}
        P_{\rm HRUS,s}(\eta_a, \eta_b, k, n) & = p_s \sum_{j=0}^{k-1} p_r^j, \\
        P_{\rm HRUS,f}(\eta_a, \eta_b, k, n) & = p_f \sum_{j=0}^{k-1} p_r^j, \\
        P_{\rm HRUS,a}(\eta_a, \eta_b, k, n) & = 1 - P_{\rm HRUS,s}(\eta_a, \eta_b, k, n) \\
        & - P_{\rm HRUS,f}(\eta_a, \eta_b, k, n).
    \end{aligned}
\end{equation*}
Note that a HRUS($k,1$) is identical to the standard RUS gate we have introduced in Section~\ref{sec_rus_gate} and a HRUS($1, n$) gate is equivalent to a boosted fusion gate~\cite{hilaire2023linear}.

Compared to RUS gates, HRUS gates (with $n\geq 2$) have a lower maximum success rate in the presence of loss than RUS gates. Yet,  
they are also faster as they usually require less trials. They should thus yield better performances when the spin coherence time is the critical figure of merit. In the main manuscript, we have assumed that a single trial of RUS or HRUS gate takes the same amount of time, which thus does not depend on the number $n$ of emitted photons per trial. This assumption holds for typically small number of photons and whenever the emission process is fast compared to the feedforward delay time between each trial.

\newpage
\bibliographystyle{unsrtnat}
\bibliography{main}

\begin{thebibliography}{107}
\providecommand{\natexlab}[1]{#1}
\providecommand{\url}[1]{\texttt{#1}}
\expandafter\ifx\csname urlstyle\endcsname\relax
  \providecommand{\doi}[1]{doi: #1}\else
  \providecommand{\doi}{doi: \begingroup \urlstyle{rm}\Url}\fi

\bibitem[Aharonov and Ben-Or(1997)]{aharonov1997fault}
Dorit Aharonov and Michael Ben-Or.
\newblock Fault-tolerant quantum computation with constant error.
\newblock In \emph{Proceedings of the twenty-ninth annual ACM symposium on
  Theory of computing}, pages 176--188, 1997.
\newblock \doi{10.48550/arXiv.quant-ph/9906129}.

\bibitem[Knill et~al.(1998)Knill, Laflamme, and Zurek]{knill1998resilient}
Emanuel Knill, Raymond Laflamme, and Wojciech~H Zurek.
\newblock Resilient quantum computation.
\newblock \emph{Science}, 279\penalty0 (5349):\penalty0 342--345, 1998.
\newblock \doi{10.1126/science.279.5349.342}.

\bibitem[Kitaev(2003)]{kitaev2003fault}
A~Yu Kitaev.
\newblock Fault-tolerant quantum computation by anyons.
\newblock \emph{Annals of physics}, 303\penalty0 (1):\penalty0 2--30, 2003.
\newblock \doi{10.1016/S0003-4916(02)00018-0}.

\bibitem[Barenco et~al.(1995)Barenco, Bennett, Cleve, DiVincenzo, Margolus,
  Shor, Sleator, Smolin, and Weinfurter]{barenco1995elementary}
Adriano Barenco, Charles~H Bennett, Richard Cleve, David~P DiVincenzo, Norman
  Margolus, Peter Shor, Tycho Sleator, John~A Smolin, and Harald Weinfurter.
\newblock Elementary gates for quantum computation.
\newblock \emph{Physical review A}, 52\penalty0 (5):\penalty0 3457, 1995.
\newblock \doi{10.1103/PhysRevA.52.3457}.

\bibitem[Raussendorf and Briegel(2001)]{raussendorf2001one}
Robert Raussendorf and Hans~J Briegel.
\newblock A one-way quantum computer.
\newblock \emph{Physical review letters}, 86\penalty0 (22):\penalty0 5188,
  2001.
\newblock \doi{10.1103/PhysRevLett.86.5188}.

\bibitem[Anders et~al.(2010)Anders, Oi, Kashefi, Browne, and
  Andersson]{anders2010ancilla}
Janet Anders, Daniel K.~L. Oi, Elham Kashefi, Dan~E Browne, and Erika
  Andersson.
\newblock Ancilla-driven universal quantum computation.
\newblock \emph{Physical Review A}, 82\penalty0 (2):\penalty0 020301(R), 2010.
\newblock \doi{10.1103/PhysRevA.82.020301}.

\bibitem[Das and Chakrabarti(2008)]{das2008colloquium}
Arnab Das and Bikas~K Chakrabarti.
\newblock Colloquium: Quantum annealing and analog quantum computation.
\newblock \emph{Reviews of Modern Physics}, 80\penalty0 (3):\penalty0 1061,
  2008.
\newblock \doi{10.1103/RevModPhys.80.1061}.

\bibitem[Van~Meter et~al.(2010)Van~Meter, Ladd, Fowler, and
  Yamamoto]{van2010distributed}
Rodney Van~Meter, Thaddeus~D Ladd, Austin~G Fowler, and Yoshihisa Yamamoto.
\newblock Distributed quantum computation architecture using semiconductor
  nanophotonics.
\newblock \emph{International Journal of Quantum Information}, 8\penalty0
  (01n02):\penalty0 295--323, 2010.
\newblock \doi{10.1142/S0219749910006435}.

\bibitem[Monroe et~al.(2014)Monroe, Raussendorf, Ruthven, Brown, Maunz, Duan,
  and Kim]{monroe2014large}
Christopher Monroe, Robert Raussendorf, Alex Ruthven, Kenneth~R Brown, Peter
  Maunz, L-M Duan, and Jungsang Kim.
\newblock Large-scale modular quantum-computer architecture with atomic memory
  and photonic interconnects.
\newblock \emph{Physical Review A}, 89\penalty0 (2):\penalty0 022317, 2014.
\newblock \doi{10.1103/PhysRevA.89.022317}.

\bibitem[Nickerson et~al.(2014)Nickerson, Fitzsimons, and
  Benjamin]{nickerson2014freely}
Naomi~H Nickerson, Joseph~F Fitzsimons, and Simon~C Benjamin.
\newblock Freely scalable quantum technologies using cells of 5-to-50 qubits
  with very lossy and noisy photonic links.
\newblock \emph{Physical Review X}, 4\penalty0 (4):\penalty0 041041, 2014.
\newblock \doi{10.1103/PhysRevX.4.041041}.

\bibitem[Bombin et~al.(2021)Bombin, Kim, Litinski, Nickerson, Pant, Pastawski,
  Roberts, and Rudolph]{bombin2021interleaving}
Hector Bombin, Isaac~H Kim, Daniel Litinski, Naomi Nickerson, Mihir Pant,
  Fernando Pastawski, Sam Roberts, and Terry Rudolph.
\newblock Interleaving: Modular architectures for fault-tolerant photonic
  quantum computing.
\newblock \emph{arXiv preprint arXiv:2103.08612}, 2021.
\newblock \doi{10.48550/arXiv.2103.08612}.

\bibitem[Bourassa et~al.(2021)Bourassa, Alexander, Vasmer, Patil, Tzitrin,
  Matsuura, Su, Baragiola, Guha, Dauphinais, et~al.]{bourassa2021blueprint}
J~Eli Bourassa, Rafael~N Alexander, Michael Vasmer, Ashlesha Patil, Ilan
  Tzitrin, Takaya Matsuura, Daiqin Su, Ben~Q Baragiola, Saikat Guha, Guillaume
  Dauphinais, et~al.
\newblock Blueprint for a scalable photonic fault-tolerant quantum computer.
\newblock \emph{Quantum}, 5:\penalty0 392, 2021.
\newblock \doi{10.22331/q-2021-02-04-392}.

\bibitem[Chamberland et~al.(2022)Chamberland, Noh, Arrangoiz-Arriola, Campbell,
  Hann, Iverson, Putterman, Bohdanowicz, Flammia, Keller, Refael, Preskill,
  Jiang, Safavi-Naeini, Painter, and Brand\~ao]{chamberland2022building}
Christopher Chamberland, Kyungjoo Noh, Patricio Arrangoiz-Arriola, Earl~T.
  Campbell, Connor~T. Hann, Joseph Iverson, Harald Putterman, Thomas~C.
  Bohdanowicz, Steven~T. Flammia, Andrew Keller, Gil Refael, John Preskill,
  Liang Jiang, Amir~H. Safavi-Naeini, Oskar Painter, and Fernando~G.S.L.
  Brand\~ao.
\newblock Building a fault-tolerant quantum computer using concatenated cat
  codes.
\newblock \emph{PRX Quantum}, 3:\penalty0 010329, Feb 2022.
\newblock \doi{10.1103/PRXQuantum.3.010329}.

\bibitem[Bartolucci et~al.(2023)Bartolucci, Birchall, Bombin, Cable, Dawson,
  Gimeno-Segovia, Johnston, Kieling, Nickerson, Pant,
  et~al.]{bartolucci2023fusion}
Sara Bartolucci, Patrick Birchall, Hector Bombin, Hugo Cable, Chris Dawson,
  Mercedes Gimeno-Segovia, Eric Johnston, Konrad Kieling, Naomi Nickerson,
  Mihir Pant, et~al.
\newblock Fusion-based quantum computation.
\newblock \emph{Nature Communications}, 14\penalty0 (1):\penalty0 912, 2023.
\newblock \doi{10.1038/s41467-023-36493-1}.

\bibitem[Thomas et~al.(2022)Thomas, Ruscio, Morin, and
  Rempe]{thomas2022efficient}
Philip Thomas, Leonardo Ruscio, Olivier Morin, and Gerhard Rempe.
\newblock Efficient generation of entangled multiphoton graph states from a
  single atom.
\newblock \emph{Nature}, 608\penalty0 (7924):\penalty0 677--681, 2022.
\newblock \doi{10.1038/s41586-022-04987-5}.

\bibitem[Blinov et~al.(2004)Blinov, Moehring, Duan, and
  Monroe]{blinov2004observation}
Boris~B Blinov, David~L Moehring, L-M Duan, and Chris Monroe.
\newblock Observation of entanglement between a single trapped atom and a
  single photon.
\newblock \emph{Nature}, 428\penalty0 (6979):\penalty0 153--157, 2004.
\newblock \doi{10.1038/nature02377}.

\bibitem[Schwartz et~al.(2016)Schwartz, Cogan, Schmidgall, Don, Gantz, Kenneth,
  Lindner, and Gershoni]{schwartz2016deterministic}
Ido Schwartz, Dan Cogan, Emma~R Schmidgall, Yaroslav Don, Liron Gantz, Oded
  Kenneth, Netanel~H Lindner, and David Gershoni.
\newblock Deterministic generation of a cluster state of entangled photons.
\newblock \emph{Science}, 354\penalty0 (6311):\penalty0 434--437, 2016.
\newblock \doi{10.1126/science.aah4758}.

\bibitem[Coste et~al.(2023)Coste, Fioretto, Belabas, Wein, Hilaire,
  Frantzeskakis, Gundin, Goes, Somaschi, Morassi, et~al.]{coste2023high}
N~Coste, DA~Fioretto, N~Belabas, SC~Wein, P~Hilaire, R~Frantzeskakis, M~Gundin,
  B~Goes, N~Somaschi, M~Morassi, et~al.
\newblock High-rate entanglement between a semiconductor spin and
  indistinguishable photons.
\newblock \emph{Nature Photonics}, pages 1--6, 2023.
\newblock \doi{10.1038/s41566-023-01186-0}.

\bibitem[Cogan et~al.(2023)Cogan, Su, Kenneth, and
  Gershoni]{cogan2023deterministic}
Dan Cogan, Zu-En Su, Oded Kenneth, and David Gershoni.
\newblock Deterministic generation of indistinguishable photons in a cluster
  state.
\newblock \emph{Nature Photonics}, pages 1--6, 2023.
\newblock \doi{10.1038/s41566-022-01152-2}.

\bibitem[Meng et~al.(2023{\natexlab{a}})Meng, Chan, Nielsen, Appel, Liu, Wang,
  Bart, Wieck, Ludwig, Midolo, et~al.]{meng2023deterministic}
Yijian Meng, Ming~Lai Chan, Rasmus~B Nielsen, Martin~H Appel, Zhe Liu, Ying
  Wang, Nikolai Bart, Andreas~D Wieck, Arne Ludwig, Leonardo Midolo, et~al.
\newblock Deterministic photon source of genuine three-qubit entanglement.
\newblock \emph{arXiv preprint arXiv:2310.12038}, 2023{\natexlab{a}}.
\newblock \doi{10.48550/arXiv.2310.12038}.

\bibitem[Meng et~al.(2023{\natexlab{b}})Meng, Faurby, Chan, Sund, Liu, Wang,
  Bart, Wieck, Ludwig, Midolo, et~al.]{meng2023photonic}
Yijian Meng, Carlos~FD Faurby, Ming~Lai Chan, Patrik~I Sund, Zhe Liu, Ying
  Wang, Nikolai Bart, Andreas~D Wieck, Arne Ludwig, Leonardo Midolo, et~al.
\newblock Photonic fusion of entangled resource states from a quantum emitter.
\newblock \emph{arXiv preprint arXiv:2312.09070}, 2023{\natexlab{b}}.
\newblock \doi{10.48550/arXiv.2312.09070}.

\bibitem[Hilaire et~al.(2023{\natexlab{a}})Hilaire, Vidro, Eisenberg, and
  Economou]{hilaire2022near}
Paul Hilaire, Leonid Vidro, Hagai~S Eisenberg, and Sophia~E Economou.
\newblock Near-deterministic hybrid generation of arbitrary photonic graph
  states using a single quantum emitter and linear optics.
\newblock \emph{Quantum}, 7:\penalty0 992, 2023{\natexlab{a}}.
\newblock \doi{10.22331/q-2023-04-27-992}.

\bibitem[Sch{\"o}n et~al.(2005)Sch{\"o}n, Solano, Verstraete, Cirac, and
  Wolf]{schon2005sequential}
Christian Sch{\"o}n, Enrique Solano, Frank Verstraete, J~Ignacio Cirac, and
  Michael~M Wolf.
\newblock Sequential generation of entangled multiqubit states.
\newblock \emph{Physical review letters}, 95\penalty0 (11):\penalty0 110503,
  2005.
\newblock \doi{10.1103/PhysRevLett.95.110503}.

\bibitem[Lindner and Rudolph(2009)]{lindner2009proposal}
Netanel~H Lindner and Terry Rudolph.
\newblock Proposal for pulsed on-demand sources of photonic cluster state
  strings.
\newblock \emph{Physical review letters}, 103\penalty0 (11):\penalty0 113602,
  2009.
\newblock \doi{10.1103/PhysRevLett.103.113602}.

\bibitem[Paesani and Brown(2023)]{paesani2022high}
Stefano Paesani and Benjamin~J. Brown.
\newblock High-threshold quantum computing by fusing one-dimensional cluster
  states.
\newblock \emph{Phys. Rev. Lett.}, 131:\penalty0 120603, Sep 2023.
\newblock \doi{10.1103/PhysRevLett.131.120603}.

\bibitem[Herrera-Mart{\'\i} et~al.(2010)Herrera-Mart{\'\i}, Fowler, Jennings,
  and Rudolph]{herrera2010photonic}
David~A Herrera-Mart{\'\i}, Austin~G Fowler, David Jennings, and Terry Rudolph.
\newblock Photonic implementation for the topological cluster-state quantum
  computer.
\newblock \emph{Physical Review A}, 82\penalty0 (3):\penalty0 032332, 2010.
\newblock \doi{10.1103/PhysRevA.82.032332}.

\bibitem[Bose et~al.(1999)Bose, Knight, Plenio, and Vedral]{bose1999proposal}
S~Bose, P.L. Knight, MB~Plenio, and V~Vedral.
\newblock Proposal for teleportation of an atomic state via cavity decay.
\newblock \emph{Physical Review Letters}, 83\penalty0 (24):\penalty0 5158,
  1999.
\newblock \doi{10.1103/PhysRevLett.83.5158}.

\bibitem[Lim et~al.(2005)Lim, Beige, and Kwek]{lim2005repeat}
Yuan~Liang Lim, Almut Beige, and Leong~Chuan Kwek.
\newblock Repeat-until-success linear optics distributed quantum computing.
\newblock \emph{Physical review letters}, 95\penalty0 (3):\penalty0 030505,
  2005.
\newblock \doi{10.1103/PhysRevLett.95.030505}.

\bibitem[Heurtel et~al.(2023)Heurtel, Fyrillas, de~Gliniasty, Le~Bihan,
  Malherbe, Pailhas, Bertasi, Bourdoncle, Emeriau, Mezher,
  et~al.]{heurtel2023perceval}
Nicolas Heurtel, Andreas Fyrillas, Gr{\'e}goire de~Gliniasty, Rapha{\"e}l
  Le~Bihan, S{\'e}bastien Malherbe, Marceau Pailhas, Eric Bertasi, Boris
  Bourdoncle, Pierre-Emmanuel Emeriau, Rawad Mezher, et~al.
\newblock Perceval: A software platform for discrete variable photonic quantum
  computing.
\newblock \emph{Quantum}, 7:\penalty0 931, 2023.
\newblock \doi{10.22331/q-2023-02-21-931}.

\bibitem[Bombin and Martin-Delgado(2006)]{bombin2006topological}
Hector Bombin and Miguel~Angel Martin-Delgado.
\newblock Topological quantum distillation.
\newblock \emph{Physical review letters}, 97\penalty0 (18):\penalty0 180501,
  2006.
\newblock \doi{10.1103/PhysRevLett.97.180501}.

\bibitem[Breuckmann and Eberhardt(2021)]{breuckmann2021quantum}
Nikolas~P Breuckmann and Jens~Niklas Eberhardt.
\newblock Quantum low-density parity-check codes.
\newblock \emph{PRX Quantum}, 2\penalty0 (4):\penalty0 040101, 2021.
\newblock \doi{10.1103/PRXQuantum.2.040101}.

\bibitem[Leverrier and Z{\'e}mor(2022)]{leverrier2022quantum}
Anthony Leverrier and Gilles Z{\'e}mor.
\newblock Quantum tanner codes.
\newblock In \emph{2022 IEEE 63rd Annual Symposium on Foundations of Computer
  Science (FOCS)}, pages 872--883. IEEE, 2022.
\newblock \doi{10.1109/FOCS54457.2022.00117}.

\bibitem[Panteleev and Kalachev(2022)]{panteleev2022asymptotically}
Pavel Panteleev and Gleb Kalachev.
\newblock Asymptotically good quantum and locally testable classical ldpc
  codes.
\newblock In \emph{Proceedings of the 54th Annual ACM SIGACT Symposium on
  Theory of Computing}, pages 375--388, 2022.
\newblock \doi{10.1145/3519935.3520017}.

\bibitem[Litinski(2019)]{litinski2019game}
Daniel Litinski.
\newblock A game of surface codes: Large-scale quantum computing with lattice
  surgery.
\newblock \emph{Quantum}, 3:\penalty0 128, 2019.
\newblock \doi{10.22331/q-2019-03-05-128}.

\bibitem[Bravyi and Kitaev(2005)]{bravyi2005universal}
Sergey Bravyi and Alexei Kitaev.
\newblock Universal quantum computation with ideal clifford gates and noisy
  ancillas.
\newblock \emph{Physical Review A}, 71\penalty0 (2):\penalty0 022316, 2005.
\newblock \doi{10.1103/PhysRevA.71.022316}.

\bibitem[Horsman et~al.(2012)Horsman, Fowler, Devitt, and
  Van~Meter]{horsman2012surface}
Clare Horsman, Austin~G Fowler, Simon Devitt, and Rodney Van~Meter.
\newblock Surface code quantum computing by lattice surgery.
\newblock \emph{New Journal of Physics}, 14\penalty0 (12):\penalty0 123011,
  2012.
\newblock \doi{10.1088/1367-2630/14/12/123011}.

\bibitem[Anderson et~al.(2014)Anderson, Duclos-Cianci, and
  Poulin]{anderson2014fault}
Jonas~T Anderson, Guillaume Duclos-Cianci, and David Poulin.
\newblock Fault-tolerant conversion between the steane and reed-muller quantum
  codes.
\newblock \emph{Physical review letters}, 113\penalty0 (8):\penalty0 080501,
  2014.
\newblock \doi{10.1103/PhysRevLett.113.080501}.

\bibitem[Bacon(2006)]{bacon2006operator}
Dave Bacon.
\newblock Operator quantum error-correcting subsystems for self-correcting
  quantum memories.
\newblock \emph{Physical Review A}, 73\penalty0 (1):\penalty0 012340, 2006.
\newblock \doi{10.1103/PhysRevA.73.012340}.

\bibitem[Hastings and Haah(2021)]{hastings2021dynamically}
Matthew~B Hastings and Jeongwan Haah.
\newblock Dynamically generated logical qubits.
\newblock \emph{Quantum}, 5:\penalty0 564, 2021.
\newblock \doi{10.22331/q-2021-10-19-564}.

\bibitem[Townsend-Teague et~al.(2023)Townsend-Teague, de~la Fuente, and
  Kesselring]{townsend2023floquetifying}
Alex Townsend-Teague, Julio~Magdalena de~la Fuente, and Markus Kesselring.
\newblock Floquetifying the colour code.
\newblock \emph{arXiv preprint arXiv:2307.11136}, 2023.
\newblock \doi{10.4204/EPTCS.384.14}.

\bibitem[Bomb{\'\i}n and Martin-Delgado(2007)]{bombin2007optimal}
H{\'e}ctor Bomb{\'\i}n and Miguel~A Martin-Delgado.
\newblock Optimal resources for topological two-dimensional stabilizer codes:
  Comparative study.
\newblock \emph{Physical Review A}, 76\penalty0 (1):\penalty0 012305, 2007.
\newblock \doi{10.1103/PhysRevA.76.012305}.

\bibitem[Connolly et~al.(2022)Connolly, Londe, Leverrier, and
  Delfosse]{connolly2022fast}
Nicholas Connolly, Vivien Londe, Anthony Leverrier, and Nicolas Delfosse.
\newblock Fast erasure decoder for a class of quantum ldpc codes.
\newblock \emph{arXiv preprint arXiv:2208.01002}, 2022.
\newblock \doi{10.48550/arXiv.2208.01002}.

\bibitem[Delfosse et~al.(2022)Delfosse, Londe, and
  Beverland]{delfosse2022toward}
Nicolas Delfosse, Vivien Londe, and Michael~E Beverland.
\newblock Toward a union-find decoder for quantum ldpc codes.
\newblock \emph{IEEE Transactions on Information Theory}, 68\penalty0
  (5):\penalty0 3187--3199, 2022.
\newblock \doi{10.1109/TIT.2022.3143452}.

\bibitem[Appel et~al.(2022)Appel, Tiranov, Pabst, Chan, Starup, Wang, Midolo,
  Tiurev, Scholz, Wieck, Ludwig, S\o{}rensen, and Lodahl]{appel2022entangling}
Martin~Hayhurst Appel, Alexey Tiranov, Simon Pabst, Ming~Lai Chan, Christian
  Starup, Ying Wang, Leonardo Midolo, Konstantin Tiurev, Sven Scholz,
  Andreas~D. Wieck, Arne Ludwig, Anders~S\o{}ndberg S\o{}rensen, and Peter
  Lodahl.
\newblock Entangling a hole spin with a time-bin photon: A waveguide approach
  for quantum dot sources of multiphoton entanglement.
\newblock \emph{Phys. Rev. Lett.}, 128:\penalty0 233602, Jun 2022.
\newblock \doi{10.1103/PhysRevLett.128.233602}.

\bibitem[Thomas et~al.(2021)Thomas, Billard, Coste, Wein, Priya, Ollivier,
  Krebs, Taza{\"\i}rt, Harouri, Lemaitre, Sagnes, L., N., J.C., and
  P.]{thomas2021bright}
S.~E. Thomas, M~Billard, N~Coste, S.~C. Wein, Priya, H~Ollivier, O~Krebs,
  L~Taza{\"\i}rt, A~Harouri, A~Lemaitre, I~Sagnes, Anton C.and~Lanco L.,
  Somaschi N., Loredo J.C., and Senellart P.
\newblock Bright polarized single-photon source based on a linear dipole.
\newblock \emph{Physical review letters}, 126\penalty0 (23):\penalty0 233601,
  2021.
\newblock \doi{10.1103/PhysRevLett.126.233601}.

\bibitem[Tomm et~al.(2021)Tomm, Javadi, Antoniadis, Najer, L{\"o}bl, Korsch,
  Schott, Valentin, Wieck, Ludwig, et~al.]{tomm2021bright}
Natasha Tomm, Alisa Javadi, Nadia~Olympia Antoniadis, Daniel Najer,
  Matthias~Christian L{\"o}bl, Alexander~Rolf Korsch, R{\"u}diger Schott,
  Sascha~Ren{\'e} Valentin, Andreas~Dirk Wieck, Arne Ludwig, et~al.
\newblock A bright and fast source of coherent single photons.
\newblock \emph{Nature Nanotechnology}, 16\penalty0 (4):\penalty0 399--403,
  2021.
\newblock \doi{10.1038/s41565-020-00831-x}.

\bibitem[Uppu et~al.(2020)Uppu, Pedersen, Wang, Olesen, Papon, Zhou, Midolo,
  Scholz, Wieck, Ludwig, et~al.]{uppu2020scalable}
Ravitej Uppu, Freja~T Pedersen, Ying Wang, Cecilie~T Olesen, Camille Papon,
  Xiaoyan Zhou, Leonardo Midolo, Sven Scholz, Andreas~D Wieck, Arne Ludwig,
  et~al.
\newblock Scalable integrated single-photon source.
\newblock \emph{Science advances}, 6\penalty0 (50):\penalty0 eabc8268, 2020.
\newblock \doi{10.1126/sciadv.abc8268}.

\bibitem[Bhaskar et~al.(2020)Bhaskar, Riedinger, Machielse, Levonian, Nguyen,
  Knall, Park, Englund, Lon{\v{c}}ar, Sukachev,
  et~al.]{bhaskar2020experimental}
Mihir~K Bhaskar, Ralf Riedinger, Bartholomeus Machielse, David~S Levonian,
  Christian~T Nguyen, Erik~N Knall, Hongkun Park, Dirk Englund, Marko
  Lon{\v{c}}ar, Denis~D Sukachev, et~al.
\newblock Experimental demonstration of memory-enhanced quantum communication.
\newblock \emph{Nature}, 580\penalty0 (7801):\penalty0 60--64, 2020.
\newblock \doi{10.1038/s41586-020-2103-5}.

\bibitem[Lenzini et~al.(2017)Lenzini, Haylock, Loredo, Abrahao, Zakaria,
  Kasture, Sagnes, Lemaitre, Phan, Dao, et~al.]{lenzini2017active}
Francesco Lenzini, Ben Haylock, Juan~C Loredo, Raphael~A Abrahao, Nor~A
  Zakaria, Sachin Kasture, Isabelle Sagnes, Aristide Lemaitre, Hoang-Phuong
  Phan, Dzung~Viet Dao, et~al.
\newblock Active demultiplexing of single photons from a solid-state source.
\newblock \emph{Laser \& Photonics Reviews}, 11\penalty0 (3):\penalty0 1600297,
  2017.
\newblock \doi{10.1002/lpor.201600297}.

\bibitem[Pont et~al.(2022)Pont, Albiero, Thomas, Spagnolo, Ceccarelli,
  Corrielli, Brieussel, Somaschi, Huet, Harouri, Lema\^{\i}tre, Sagnes,
  Belabas, Sciarrino, Osellame, Senellart, and Crespi]{pont2022quantifying}
Mathias Pont, Riccardo Albiero, Sarah~E. Thomas, Nicol\`o Spagnolo, Francesco
  Ceccarelli, Giacomo Corrielli, Alexandre Brieussel, Niccolo Somaschi, H\^elio
  Huet, Abdelmounaim Harouri, Aristide Lema\^{\i}tre, Isabelle Sagnes, Nadia
  Belabas, Fabio Sciarrino, Roberto Osellame, Pascale Senellart, and Andrea
  Crespi.
\newblock Quantifying $n$-photon indistinguishability with a cyclic integrated
  interferometer.
\newblock \emph{Phys. Rev. X}, 12:\penalty0 031033, Sep 2022.
\newblock \doi{10.1103/PhysRevX.12.031033}.

\bibitem[Hansen et~al.(2023)Hansen, Carosini, Jehle, Giorgino, Houvenaghel,
  Vyvlecka, Loredo, and Walther]{hansen2023single}
Lena~M Hansen, Lorenzo Carosini, Lennart Jehle, Francesco Giorgino, Romane
  Houvenaghel, Michal Vyvlecka, Juan~C Loredo, and Philip Walther.
\newblock Single-active-element demultiplexed multi-photon source.
\newblock \emph{arXiv preprint arXiv:2304.12956}, 2023.
\newblock \doi{10.1364/OPTICAQ.494643}.

\bibitem[Moussa(2016)]{moussa2016transversal}
Jonathan~E Moussa.
\newblock Transversal clifford gates on folded surface codes.
\newblock \emph{Physical Review A}, 94\penalty0 (4):\penalty0 042316, 2016.
\newblock \doi{10.1103/PhysRevA.94.042316}.

\bibitem[Campbell(2019)]{campbell2019theory}
Earl~T Campbell.
\newblock A theory of single-shot error correction for adversarial noise.
\newblock \emph{Quantum Science and Technology}, 4\penalty0 (2):\penalty0
  025006, 2019.
\newblock \doi{10.1088/2058-9565/aafc8f}.

\bibitem[Litinski and Nickerson(2022)]{litinski2022active}
Daniel Litinski and Naomi Nickerson.
\newblock Active volume: An architecture for efficient fault-tolerant quantum
  computers with limited non-local connections.
\newblock \emph{arXiv preprint arXiv:2211.15465}, 2022.
\newblock \doi{10.48550/arXiv.2211.15465}.

\bibitem[Paetznick and Svore(2013)]{paetznick2013repeat}
Adam Paetznick and Krysta~M Svore.
\newblock Repeat-until-success: Non-deterministic decomposition of single-qubit
  unitaries.
\newblock \emph{arXiv preprint arXiv:1311.1074}, 2013.
\newblock \doi{10.48550/arXiv.1311.1074}.

\bibitem[Shah and Oi(2013)]{shah2013ancilla}
Kerem~Halil Shah and Daniel Kuan~Li Oi.
\newblock Ancilla driven quantum computation with arbitrary entangling
  strength.
\newblock \emph{arXiv preprint arXiv:1303.2066}, 2013.
\newblock \doi{10.48550/arXiv.1303.2066}.

\bibitem[Lim et~al.(2006)Lim, Barrett, Beige, Kok, and Kwek]{lim2006repeat}
Yuan~Liang Lim, Sean~D Barrett, Almut Beige, Pieter Kok, and Leong~Chuan Kwek.
\newblock Repeat-until-success quantum computing using stationary and flying
  qubits.
\newblock \emph{Physical Review A}, 73\penalty0 (1):\penalty0 012304, 2006.
\newblock \doi{10.1103/PhysRevA.73.012304}.

\bibitem[Raussendorf et~al.(2007)Raussendorf, Harrington, and
  Goyal]{raussendorf2007topological}
Robert Raussendorf, Jim Harrington, and Kovid Goyal.
\newblock Topological fault-tolerance in cluster state quantum computation.
\newblock \emph{New Journal of Physics}, 9\penalty0 (6):\penalty0 199, 2007.
\newblock \doi{10.1088/1367-2630/9/6/199}.

\bibitem[Li et~al.(2010)Li, Barrett, Stace, and Benjamin]{li2010fault}
Ying Li, Sean~D Barrett, Thomas~M Stace, and Simon~C Benjamin.
\newblock Fault tolerant quantum computation with nondeterministic gates.
\newblock \emph{Physical review letters}, 105\penalty0 (25):\penalty0 250502,
  2010.
\newblock \doi{10.1103/PhysRevLett.105.250502}.

\bibitem[Bell et~al.(2022)Bell, Bulmer, Jones, Paesani, McCutcheon, and
  Laing]{bell2022protocol}
Thomas~J Bell, Jacob~FF Bulmer, Alex~E Jones, Stefano Paesani, Dara~PS
  McCutcheon, and Anthony Laing.
\newblock Protocol for generation of high-dimensional entanglement from an
  array of non-interacting photon emitters.
\newblock \emph{New Journal of Physics}, 24\penalty0 (1):\penalty0 013032,
  2022.
\newblock \doi{10.1088/1367-2630/ac475d}.

\bibitem[Lee et~al.(2019)Lee, Villa, Bennett, Stevenson, Ellis, Farrer,
  Ritchie, and Shields]{lee2019quantum}
JP~Lee, B~Villa, AJ~Bennett, RM~Stevenson, DJP Ellis, I~Farrer, DA~Ritchie, and
  AJ~Shields.
\newblock A quantum dot as a source of time-bin entangled multi-photon states.
\newblock \emph{Quantum Science and Technology}, 4\penalty0 (2):\penalty0
  025011, 2019.
\newblock \doi{10.1088/2058-9565/ab0a9b}.

\bibitem[Vezvaee et~al.(2022)Vezvaee, Hilaire, Doty, and
  Economou]{vezvaee2022deterministic}
Arian Vezvaee, Paul Hilaire, Matthew~F Doty, and Sophia~E Economou.
\newblock Deterministic generation of entangled photonic cluster states from
  quantum dot molecules.
\newblock \emph{Physical Review Applied}, 18\penalty0 (6):\penalty0 L061003,
  2022.
\newblock \doi{10.1103/PhysRevApplied.18.L061003}.

\bibitem[Tiurev et~al.(2022)Tiurev, Appel, Mirambell, Lauritzen, Tiranov,
  Lodahl, and S{\o}rensen]{tiurev2022high}
Konstantin Tiurev, Martin~Hayhurst Appel, Pol~Llopart Mirambell, Mikkel~Bloch
  Lauritzen, Alexey Tiranov, Peter Lodahl, and Anders~S{\o}ndberg S{\o}rensen.
\newblock High-fidelity multiphoton-entangled cluster state with solid-state
  quantum emitters in photonic nanostructures.
\newblock \emph{Physical Review A}, 105\penalty0 (3):\penalty0 L030601, 2022.
\newblock \doi{10.1103/PhysRevA.105.L030601}.

\bibitem[Vasconcelos et~al.(2020)Vasconcelos, Reisenbauer, Salter, Wachter,
  Wirtitsch, Schmiedmayer, Walther, and Trupke]{vasconcelos2020scalable}
Rui Vasconcelos, Sarah Reisenbauer, Cameron Salter, Georg Wachter, Daniel
  Wirtitsch, J{\"o}rg Schmiedmayer, Philip Walther, and Michael Trupke.
\newblock Scalable spin--photon entanglement by time-to-polarization
  conversion.
\newblock \emph{npj Quantum Information}, 6\penalty0 (1):\penalty0 9, 2020.
\newblock \doi{10.1038/s41534-019-0236-x}.

\bibitem[Besse et~al.(2020)Besse, Reuer, Collodo, Wulff, Wernli, Copetudo,
  Malz, Magnard, Akin, Gabureac, et~al.]{besse2020realizing}
Jean-Claude Besse, Kevin Reuer, Michele~C Collodo, Arne Wulff, Lucien Wernli,
  Adrian Copetudo, Daniel Malz, Paul Magnard, Abdulkadir Akin, Mihai Gabureac,
  et~al.
\newblock Realizing a deterministic source of multipartite-entangled photonic
  qubits.
\newblock \emph{Nature communications}, 11\penalty0 (1):\penalty0 4877, 2020.
\newblock \doi{10.1038/s41467-020-18635-x}.

\bibitem[Liu et~al.(2022)Liu, Barnes, and Economou]{liu2022proposal}
Chenxu Liu, Edwin Barnes, and Sophia Economou.
\newblock Proposal for generating complex microwave graph states using
  superconducting circuits.
\newblock In \emph{Quantum 2.0}, pages QTu2A--21. Optica Publishing Group,
  2022.
\newblock \doi{10.1364/QUANTUM.2022.QTu2A.21}.

\bibitem[Gidney(2021)]{gidney2021stim}
Craig Gidney.
\newblock Stim: a fast stabilizer circuit simulator.
\newblock \emph{Quantum}, 5:\penalty0 497, 2021.
\newblock \doi{10.22331/q-2021-07-06-497}.

\bibitem[Higgott(2022)]{higgott2022pymatching}
Oscar Higgott.
\newblock Pymatching: A python package for decoding quantum codes with
  minimum-weight perfect matching.
\newblock \emph{ACM Transactions on Quantum Computing}, 3\penalty0
  (3):\penalty0 1--16, 2022.
\newblock \doi{10.1145/3505637}.

\bibitem[Higgott and Gidney(2023)]{higgott2023sparse}
Oscar Higgott and Craig Gidney.
\newblock Sparse blossom: correcting a million errors per core second with
  minimum-weight matching.
\newblock \emph{arXiv preprint arXiv:2303.15933}, 2023.
\newblock \doi{10.48550/arXiv.2303.15933}.

\bibitem[Stace and Barrett(2010)]{stace2010error}
Thomas~M Stace and Sean~D Barrett.
\newblock Error correction and degeneracy in surface codes suffering loss.
\newblock \emph{Physical Review A}, 81\penalty0 (2):\penalty0 022317, 2010.
\newblock \doi{10.1103/PhysRevA.81.022317}.

\bibitem[Hilaire et~al.(2023{\natexlab{b}})Hilaire, Castor, Barnes, Economou,
  and Grosshans]{hilaire2023linear}
Paul Hilaire, Yaron Castor, Edwin Barnes, Sophia~E. Economou, and Fr\'ed\'eric
  Grosshans.
\newblock Linear optical logical bell state measurements with optimal
  loss-tolerance threshold.
\newblock \emph{PRX Quantum}, 4:\penalty0 040322, Nov 2023{\natexlab{b}}.
\newblock \doi{10.1103/PRXQuantum.4.040322}.

\bibitem[Roffe et~al.(2020)Roffe, White, Burton, and
  Campbell]{roffe2020decoding}
Joschka Roffe, David~R White, Simon Burton, and Earl Campbell.
\newblock Decoding across the quantum low-density parity-check code landscape.
\newblock \emph{Physical Review Research}, 2\penalty0 (4):\penalty0 043423,
  2020.
\newblock \doi{10.1103/PhysRevResearch.2.043423}.

\bibitem[Tillich and Z{\'e}mor(2013)]{tillich2013quantum}
Jean-Pierre Tillich and Gilles Z{\'e}mor.
\newblock Quantum ldpc codes with positive rate and minimum distance
  proportional to the square root of the blocklength.
\newblock \emph{IEEE Transactions on Information Theory}, 60\penalty0
  (2):\penalty0 1193--1202, 2013.
\newblock \doi{10.1109/TIT.2013.2292061}.

\bibitem[Bravyi et~al.(2023)Bravyi, Cross, Gambetta, Maslov, Rall, and
  Yoder]{bravyi2023high}
Sergey Bravyi, Andrew~W Cross, Jay~M Gambetta, Dmitri Maslov, Patrick Rall, and
  Theodore~J Yoder.
\newblock High-threshold and low-overhead fault-tolerant quantum memory.
\newblock \emph{arXiv preprint arXiv:2308.07915}, 2023.
\newblock \doi{10.1038/s41586-024-07107-7}.

\bibitem[Breuckmann and Londe(2021)]{breuckmann2021single}
Nikolas~P Breuckmann and Vivien Londe.
\newblock Single-shot decoding of linear rate ldpc quantum codes with high
  performance.
\newblock \emph{IEEE Transactions on Information Theory}, 68\penalty0
  (1):\penalty0 272--286, 2021.
\newblock \doi{10.1109/TIT.2021.3122352}.

\bibitem[Xu et~al.(2022)Xu, Seif, Yan, Mannucci, Sane, Van~Meter, Cleland, and
  Jiang]{xu2022distributed}
Qian Xu, Alireza Seif, Haoxiong Yan, Nam Mannucci, Bernard~Ousmane Sane, Rodney
  Van~Meter, Andrew~N Cleland, and Liang Jiang.
\newblock Distributed quantum error correction for chip-level catastrophic
  errors.
\newblock \emph{Physical review letters}, 129\penalty0 (24):\penalty0 240502,
  2022.
\newblock \doi{10.1103/PhysRevLett.129.240502}.

\bibitem[Ang et~al.(2022)Ang, Carini, Chen, Chuang, DeMarco, Economou,
  Eickbusch, Faraon, Fu, Girvin, et~al.]{ang2022architectures}
James Ang, Gabriella Carini, Yanzhu Chen, Isaac Chuang, Michael~Austin DeMarco,
  Sophia~E Economou, Alec Eickbusch, Andrei Faraon, Kai-Mei Fu, Steven~M
  Girvin, et~al.
\newblock Architectures for multinode superconducting quantum computers.
\newblock \emph{arXiv preprint arXiv:2212.06167}, 2022.
\newblock \doi{10.48550/arXiv.2212.06167}.

\bibitem[L{\"o}bl et~al.(2024)L{\"o}bl, Paesani, and S{\o}rensen]{lobl2023loss}
Matthias~C L{\"o}bl, Stefano Paesani, and Anders~S S{\o}rensen.
\newblock Loss-tolerant architecture for quantum computing with quantum
  emitters.
\newblock \emph{Quantum}, 8:\penalty0 1302, 2024.
\newblock \doi{10.22331/q-2024-03-28-1302}.

\bibitem[Omkar et~al.(2022)Omkar, Lee, Teo, Lee, and Jeong]{omkar2022all}
Srikrishna Omkar, Seok-Hyung Lee, Yong~Siah Teo, Seung-Woo Lee, and Hyunseok
  Jeong.
\newblock All-photonic architecture for scalable quantum computing with
  greenberger-horne-zeilinger states.
\newblock \emph{PRX Quantum}, 3\penalty0 (3):\penalty0 030309, 2022.
\newblock \doi{10.1103/PRXQuantum.3.030309}.

\bibitem[Pankovich et~al.(2023)Pankovich, Kan, Wan, Ostmann, Neville, Omkar,
  Sohbi, and Br\'adler]{pankovich2023high}
Brendan Pankovich, Angus Kan, Kwok~Ho Wan, Maike Ostmann, Alex Neville,
  Srikrishna Omkar, Adel Sohbi, and Kamil Br\'adler.
\newblock High photon-loss threshold quantum computing using ghz-state
  measurements.
\newblock \emph{arXiv preprint arXiv:2308.04192}, 2023.
\newblock \doi{10.48550/arXiv.2308.04192}.

\bibitem[Li et~al.(2022)Li, Economou, and Barnes]{li2022photonic}
Bikun Li, Sophia~E Economou, and Edwin Barnes.
\newblock Photonic resource state generation from a minimal number of quantum
  emitters.
\newblock \emph{npj Quantum Information}, 8\penalty0 (1):\penalty0 11, 2022.
\newblock \doi{10.1038/s41534-022-00522-6}.

\bibitem[Maring et~al.(2024)Maring, Fyrillas, Pont, Ivanov, Stepanov, Margaria,
  Hease, Pishchagin, Lema{\^\i}tre, Sagnes, et~al.]{maring2023general}
Nicolas Maring, Andreas Fyrillas, Mathias Pont, Edouard Ivanov, Petr Stepanov,
  Nico Margaria, William Hease, Anton Pishchagin, Aristide Lema{\^\i}tre,
  Isabelle Sagnes, et~al.
\newblock A versatile single-photon-based quantum computing platform.
\newblock \emph{Nature Photonics}, 18\penalty0 (6):\penalty0 603--609, 2024.
\newblock \doi{10.1038/s41566-024-01403-4}.

\bibitem[Atat{\"u}re et~al.(2018)Atat{\"u}re, Englund, Vamivakas, Lee, and
  Wrachtrup]{atature2018material}
Mete Atat{\"u}re, Dirk Englund, Nick Vamivakas, Sang-Yun Lee, and Joerg
  Wrachtrup.
\newblock Material platforms for spin-based photonic quantum technologies.
\newblock \emph{Nature Reviews Materials}, 3\penalty0 (5):\penalty0 38--51,
  2018.
\newblock \doi{10.1038/s41578-018-0008-9}.

\bibitem[Bock et~al.(2018)Bock, Eich, Kucera, Kreis, Lenhard, Becher, and
  Eschner]{bock2018high}
Matthias Bock, Pascal Eich, Stephan Kucera, Matthias Kreis, Andreas Lenhard,
  Christoph Becher, and J{\"u}rgen Eschner.
\newblock High-fidelity entanglement between a trapped ion and a telecom photon
  via quantum frequency conversion.
\newblock \emph{Nature communications}, 9\penalty0 (1):\penalty0 1998, 2018.
\newblock \doi{10.1038/s41467-018-04341-2}.

\bibitem[Tchebotareva et~al.(2019)Tchebotareva, Hermans, Humphreys, Voigt,
  Harmsma, Cheng, Verlaan, Dijkhuizen, De~Jong, Dr{\'e}au, and
  R.]{tchebotareva2019entanglement}
Anna Tchebotareva, Sophie L.~N. Hermans, Peter~C Humphreys, Dirk Voigt, Peter~J
  Harmsma, Lun~K Cheng, Ad~L Verlaan, Niels Dijkhuizen, Wim De~Jong, Ana{\"\i}s
  Dr{\'e}au, and Hanson R.
\newblock Entanglement between a diamond spin qubit and a photonic time-bin
  qubit at telecom wavelength.
\newblock \emph{Physical review letters}, 123\penalty0 (6):\penalty0 063601,
  2019.
\newblock \doi{10.1103/PhysRevLett.123.063601}.

\bibitem[Siyushev et~al.(2014)Siyushev, Xia, Reuter, Jamali, Zhao, Yang, Duan,
  Kukharchyk, Wieck, Kolesov, et~al.]{siyushev2014coherent}
P~Siyushev, K~Xia, R~Reuter, M~Jamali, N~Zhao, N~Yang, C~Duan, N~Kukharchyk,
  AD~Wieck, R~Kolesov, et~al.
\newblock Coherent properties of single rare-earth spin qubits.
\newblock \emph{Nature communications}, 5\penalty0 (1):\penalty0 3895, 2014.
\newblock \doi{10.1038/ncomms4895}.

\bibitem[Ruskuc et~al.(2022)Ruskuc, Wu, Rochman, Choi, and
  Faraon]{ruskuc2022nuclear}
Andrei Ruskuc, Chun-Ju Wu, Jake Rochman, Joonhee Choi, and Andrei Faraon.
\newblock Nuclear spin-wave quantum register for a solid-state qubit.
\newblock \emph{Nature}, 602\penalty0 (7897):\penalty0 408--413, 2022.
\newblock \doi{10.1038/s41586-021-04293-6}.

\bibitem[Higginbottom et~al.(2022)Higginbottom, Kurkjian, Chartrand, Kazemi,
  Brunelle, MacQuarrie, Klein, Lee-Hone, Stacho, Ruether,
  et~al.]{higginbottom2022optical}
Daniel~B Higginbottom, Alexander~TK Kurkjian, Camille Chartrand, Moein Kazemi,
  Nicholas~A Brunelle, Evan~R MacQuarrie, James~R Klein, Nicholas~R Lee-Hone,
  Jakub Stacho, Myles Ruether, et~al.
\newblock Optical observation of single spins in silicon.
\newblock \emph{Nature}, 607\penalty0 (7918):\penalty0 266--270, 2022.
\newblock \doi{10.1038/s41586-022-04821-y}.

\bibitem[Lee et~al.(2023)Lee, Islam, Harper, Buyukkaya, Higginbottom, Simmons,
  and Waks]{lee2023high}
Chang-Min Lee, Fariba Islam, Samuel Harper, Mustafa~Atabey Buyukkaya, Daniel
  Higginbottom, Stephanie Simmons, and Edo Waks.
\newblock High-efficiency single photon emission from a silicon t-center in a
  nanobeam.
\newblock \emph{ACS Photonics}, 10\penalty0 (11):\penalty0 3844--3849, 2023.
\newblock \doi{10.1021/acsphotonics.3c01142}.

\bibitem[Castelletto et~al.(2020)Castelletto, Inam, Sato, and
  Boretti]{castelletto2020hexagonal}
Stefania Castelletto, Faraz~A Inam, Shin-ichiro Sato, and Alberto Boretti.
\newblock Hexagonal boron nitride: a review of the emerging material platform
  for single-photon sources and the spin--photon interface.
\newblock \emph{Beilstein Journal of Nanotechnology}, 11\penalty0 (1):\penalty0
  740--769, 2020.
\newblock \doi{10.3762/bjnano.11.61}.

\bibitem[Kubanek(2022)]{kubanek2022coherent}
Alexander Kubanek.
\newblock Coherent quantum emitters in hexagonal boron nitride.
\newblock \emph{Advanced Quantum Technologies}, 5\penalty0 (9):\penalty0
  2200009, 2022.
\newblock \doi{10.1002/qute.202200009}.

\bibitem[Patel et~al.(2024)Patel, Fishman, Huang, Gusdorff, Fehr, Hopper,
  Breitweiser, Porat, Flatt{\'e}, and Bassett]{patel2023dynamical}
Raj~N Patel, Rebecca~EK Fishman, Tzu-Yung Huang, Jordan~A Gusdorff, David~A
  Fehr, David~A Hopper, S~Alex Breitweiser, Benjamin Porat, Michael~E
  Flatt{\'e}, and Lee~C Bassett.
\newblock Room temperature dynamics of an optically addressable single spin in
  hexagonal boron nitride.
\newblock \emph{Nano Letters}, 2024.
\newblock \doi{10.1021/acs.nanolett.4c01333}.

\bibitem[Auger et~al.(2018)Auger, Anwar, Gimeno-Segovia, Stace, and
  Browne]{auger2018fault}
James~M Auger, Hussain Anwar, Mercedes Gimeno-Segovia, Thomas~M Stace, and
  Dan~E Browne.
\newblock Fault-tolerant quantum computation with nondeterministic entangling
  gates.
\newblock \emph{Physical Review A}, 97\penalty0 (3):\penalty0 030301(R), 2018.
\newblock \doi{10.1103/PhysRevA.97.030301}.

\bibitem[Bell et~al.(2023)Bell, Pettersson, and Paesani]{bell2022optimising}
Thomas~J Bell, Love~A Pettersson, and Stefano Paesani.
\newblock Optimizing graph codes for measurement-based loss tolerance.
\newblock \emph{PRX Quantum}, 4\penalty0 (2):\penalty0 020328, 2023.
\newblock \doi{10.1103/PRXQuantum.4.020328}.

\bibitem[Bomb{\'\i}n et~al.(2023)Bomb{\'\i}n, Dawson, Nickerson, Pant, and
  Sullivan]{bombin2023increasing}
Hector Bomb{\'\i}n, Chris Dawson, Naomi Nickerson, Mihir Pant, and Jordan
  Sullivan.
\newblock Increasing error tolerance in quantum computers with dynamic bias
  arrangement.
\newblock \emph{arXiv preprint arXiv:2303.16122}, 2023.
\newblock \doi{10.48550/arXiv.2303.16122}.

\bibitem[Gottesman(2013)]{gottesman2013fault}
Daniel Gottesman.
\newblock Fault-tolerant quantum computation with constant overhead.
\newblock \emph{arXiv preprint arXiv:1310.2984}, 2013.
\newblock \doi{10.48550/arXiv.1310.2984}.

\bibitem[Fawzi et~al.(2020)Fawzi, Grospellier, and
  Leverrier]{fawzi2020constant}
Omar Fawzi, Antoine Grospellier, and Anthony Leverrier.
\newblock Constant overhead quantum fault tolerance with quantum expander
  codes.
\newblock \emph{Communications of the ACM}, 64\penalty0 (1):\penalty0 106--114,
  2020.
\newblock \doi{10.1145/3434163}.

\bibitem[Cohen et~al.(2022)Cohen, Kim, Bartlett, and Brown]{cohen2022low}
Lawrence~Z Cohen, Isaac~H Kim, Stephen~D Bartlett, and Benjamin~J Brown.
\newblock Low-overhead fault-tolerant quantum computing using long-range
  connectivity.
\newblock \emph{Science Advances}, 8\penalty0 (20):\penalty0 eabn1717, 2022.
\newblock \doi{10.1126/sciadv.abn1717}.

\bibitem[Breuer and Petruccione(2002)]{breuer2002theory}
Heinz-Peter Breuer and Francesco Petruccione.
\newblock \emph{The theory of open quantum systems}.
\newblock Oxford University Press, USA, 2002.
\newblock \doi{10.1093/acprof:oso/9780199213900.001.0001}.

\bibitem[Hong et~al.(1987)Hong, Ou, and Mandel]{hong1987measurement}
Chong-Ki Hong, Zhe-Yu Ou, and Leonard Mandel.
\newblock Measurement of subpicosecond time intervals between two photons by
  interference.
\newblock \emph{Physical review letters}, 59\penalty0 (18):\penalty0 2044,
  1987.
\newblock \doi{10.1103/PhysRevLett.59.2044}.

\bibitem[Zaporski et~al.(2023)Zaporski, Shofer, Bodey, Manna, Gillard, Appel,
  Schimpf, Covre~da Silva, Jarman, Delamare, et~al.]{zaporski2023ideal}
Leon Zaporski, Noah Shofer, Jonathan~H Bodey, Santanu Manna, George Gillard,
  Martin~Hayhurst Appel, Christian Schimpf, Saimon~Filipe Covre~da Silva, John
  Jarman, Geoffroy Delamare, et~al.
\newblock Ideal refocusing of an optically active spin qubit under strong
  hyperfine interactions.
\newblock \emph{Nature Nanotechnology}, 18\penalty0 (3):\penalty0 257--263,
  2023.
\newblock \doi{10.1038/s41565-022-01282-2}.

\bibitem[Claudon et~al.(2010)Claudon, Bleuse, Malik, Bazin, Jaffrennou,
  Gregersen, Sauvan, Lalanne, and G{\'e}rard]{claudon2010highly}
Julien Claudon, Jo{\"e}l Bleuse, Nitin~Singh Malik, Maela Bazin, P{\'e}rine
  Jaffrennou, Niels Gregersen, Christophe Sauvan, Philippe Lalanne, and
  Jean-Michel G{\'e}rard.
\newblock A highly efficient single-photon source based on a quantum dot in a
  photonic nanowire.
\newblock \emph{Nature Photonics}, 4\penalty0 (3):\penalty0 174--177, 2010.
\newblock \doi{10.1038/nphoton.2009.287x}.

\bibitem[Ding et~al.(2023)Ding, Guo, Xu, Liu, Zou, Zhao, Ge, Zhang, Liu, Chen,
  et~al.]{ding2023high}
Xing Ding, Yong-Peng Guo, Mo-Chi Xu, Run-Ze Liu, Geng-Yan Zou, Jun-Yi Zhao,
  Zhen-Xuan Ge, Qi-Hang Zhang, Hua-Liang Liu, Ming-Cheng Chen, et~al.
\newblock High-efficiency single-photon source above the loss-tolerant
  threshold for efficient linear optical quantum computing.
\newblock \emph{arXiv preprint arXiv:2311.08347}, 2023.
\newblock \doi{https://doi.org/10.48550/arXiv.2311.08347}.

\bibitem[Somaschi et~al.(2016)Somaschi, Giesz, De~Santis, Loredo, Almeida,
  Hornecker, Portalupi, Grange, Anton, Demory, et~al.]{somaschi2016near}
Niccolo Somaschi, Valerian Giesz, Lorenzo De~Santis, JC~Loredo, Marcelo~P
  Almeida, Gaston Hornecker, S~Luca Portalupi, Thomas Grange, Carlos Anton,
  Justin Demory, et~al.
\newblock Near-optimal single-photon sources in the solid state.
\newblock \emph{Nature Photonics}, 10\penalty0 (5):\penalty0 340--345, 2016.
\newblock \doi{10.1038/nphoton.2016.23}.

\bibitem[Zhai et~al.(2022)Zhai, Nguyen, Spinnler, Ritzmann, L{\"o}bl, Wieck,
  Ludwig, Javadi, and Warburton]{zhai2022quantum}
Liang Zhai, Giang~N Nguyen, Clemens Spinnler, Julian Ritzmann, Matthias~C
  L{\"o}bl, Andreas~D Wieck, Arne Ludwig, Alisa Javadi, and Richard~J
  Warburton.
\newblock Quantum interference of identical photons from remote gaas quantum
  dots.
\newblock \emph{Nature nanotechnology}, 17\penalty0 (8):\penalty0 829--833,
  2022.
\newblock \doi{10.1038/s41565-022-01131-2}.

\bibitem[McEwen et~al.(2023)McEwen, Bacon, and Gidney]{mcewen2023relaxing}
Matt McEwen, Dave Bacon, and Craig Gidney.
\newblock Relaxing hardware requirements for surface code circuits using
  time-dynamics.
\newblock \emph{arXiv preprint arXiv:2302.02192}, 2023.
\newblock \doi{10.22331/q-2023-11-07-1172}.

\bibitem[Suchara et~al.(2015)Suchara, Cross, and Gambetta]{Suchara2015Leakage}
Martin Suchara, Andrew~W. Cross, and Jay~M. Gambetta.
\newblock Leakage suppression in the toric code.
\newblock In \emph{2015 IEEE International Symposium on Information Theory
  (ISIT)}, pages 1119--1123, 2015.
\newblock \doi{10.1109/ISIT.2015.7282629}.

\end{thebibliography}

\end{document}